\documentclass{emulateapj}
\usepackage{natbib}
\slugcomment{ApJ, in prep.}
\shorttitle{Five New Local Volume Dwarf Galaxies Associated with UCHVCs}
\shortauthors{Sand et al.}

\begin{document}
 \title{A comprehensive archival search for counterparts to Ultra-compact High Velocity Clouds: Five Local Volume Dwarf Galaxies}

\author{D. J. Sand,$\!$\altaffilmark{1} D. Crnojevi\'{c},$\!$\altaffilmark{1} P. Bennet,$\!$\altaffilmark{1}  B. Willman,$\!$\altaffilmark{2} J. Hargis,$\!$\altaffilmark{2} J. Strader,$\!$\altaffilmark{3}   E. Olszewski,$\!$\altaffilmark{4} E.J. Tollerud,$\!$\altaffilmark{5,12} J.D. Simon,$\!$\altaffilmark{6} N. Caldwell,$\!$\altaffilmark{8} P. Guhathakurta,$\!$\altaffilmark{10} B. L. James,$\!$\altaffilmark{7} S. Koposov,$\!$\altaffilmark{7} B. McLeod,$\!$\altaffilmark{8} N. Morrell$\!$\altaffilmark{9},  M. Peacock,$\!$\altaffilmark{3} R. Salinas,$\!$\altaffilmark{3} A.C. Seth,$\!$\altaffilmark{11} D. P. Stark,$\!$\altaffilmark{4} E. Toloba$\!$\altaffilmark{1}  } \email{david.sand@ttu.edu}

\begin{abstract}

We report five Local Volume dwarf galaxies (two of which are presented here for the first time) uncovered during a comprehensive archival search for optical counterparts to ultra-compact high velocity clouds (UCHVCs).  The UCHVC population of HI clouds are thought to be candidate gas-rich, low mass halos at the edge of the Local Group and beyond, but no comprehensive search for stellar counterparts to these systems has been presented.  Careful visual inspection of all publicly available optical and ultraviolet imaging at the position of the UCHVCs revealed six blue, diffuse counterparts with a morphology consistent with a faint dwarf galaxy beyond the Local Group.  Optical spectroscopy of all six candidate dwarf counterparts show that five have an H$\alpha$-derived velocity consistent with the coincident HI cloud, confirming their association; the sixth diffuse counterpart is likely a background object.  The size and luminosity of the UCHVC dwarfs is consistent with other known Local Volume dwarf irregular galaxies.  The gas fraction ($M_{HI}/M_{star}$) of the five dwarfs are generally consistent with that of dwarf irregular galaxies in the Local Volume, although ALFALFA-Dw1 (associated with ALFALFA UCHVC HVC274.68+74.70$-$123) has a very high $M_{HI}/M_{star}$$\sim$40.  Despite the heterogenous nature of our search, we demonstrate that the current dwarf companions to UCHVCs are at the edge of detectability due to their low surface brightness, and that deeper searches are likely to find more stellar systems.  If more sensitive searches do not reveal further stellar counterparts to UCHVCs, then the dearth of such systems around the Local Group may be in conflict with $\Lambda$CDM simulations.
\end{abstract}
 
\altaffiltext{1}{Texas Tech University, Physics Department, Box 41051, Lubbock, TX 79409-1051, USA}
\altaffiltext{2}{Department of Physics, Haverford College, 370 Lancaster Avenue, Haverford, PA 19041, USA}
\altaffiltext{3}{Michigan State University, Department of Physics and Astronomy, East Lansing, MI 48824, USA}
\altaffiltext{4}{University of Arizona, Tucson, AZ 85721, USA}
\altaffiltext{5}{Yale University, Astronomy Department, PO Box 208101, New Haven, CT 06510, USA}
\altaffiltext{6}{Observatories of the Carnegie Institution for Science, 813 Santa Barbara Street, Pasadena, CA 91101, USA}
\altaffiltext{7}{Institute of Astronomy, University of Cambridge, Madingley Road, Cambridge, CB3 0HA}
\altaffiltext{8}{Harvard-Smithsonian Center for Astrophysics, Cambridge, MA 02138, USA}
\altaffiltext{9}{Las Campanas Observatory, Carnegie Observatories, Casilla 601, La Serena, Chile}
\altaffiltext{10}{UCO/Lick Observatory, University of California, Santa Cruz, 1156 High Street, Santa Cruz, CA 95064, USA}
\altaffiltext{11}{Department of Physics and Astronomy, University of Utah, Salt Lake City, UT 84112, USA}
\altaffiltext{12}{Hubble Fellow}

\section{Introduction}\label{sec:intro}

Blind HI surveys have long been used as a foundation for studies of low surface brightness (LSB) galaxies \citep[e.g.][]{zwaan97,doyle05,rosenberg05,west10,Cannon11,Haynes11}. Selecting galaxies on the basis of HI 
mitigates against the optical bias toward galaxies with high surface brightness, and allows targeted follow-up to discover new LSB galaxies \citep[e.g.][]{spitzak98}. Extragalactic HI clouds without obvious optical counterparts are themselves interesting probes of (i) our understanding of the extreme limits of galaxy formation and (ii) our census of dark matter halos in the Universe \citep[e.g.][]{Adams15,Cannon15,J15}.  Such objects may be ``dark galaxies": dark matter halos with few or no stars.




Interest in this interpretation has been stoked by tensions between our expectations and observations of the Universe on sub-galactic scales.  For example, the discovery of numerous ultra-faint dwarf galaxies around the Milky Way and M31 have underscored the vast incompleteness in our census of dwarf galaxies in even the local universe \citep[e.g.][]{Willman10,hargis14}.  This limits our ability to understand the origin of the ``missing satellite problem": the discrepancy between the small number of dwarfs observed around the Milky Way and M31 and the huge number of dark matter sub-halos predicted by theory \citep[e.g.][]{Klypin99,M99b}.


The advantage of HI galaxy selection is that it offers an opportunity to overcome this incompleteness for nearby LSB galaxies (trading the disadvantage that some LSB galaxies have little or no HI gas). HI clouds are intimately tied to the history of the ``missing satellite problem": an early solution was that a subset of the high-velocity HI clouds (HVCs) observed in the Local Group could inhabit some of the ``missing" dark matter halos \citep{braun99,blitz99}. These original studies focused on the HVC catalog of \citet{wakker91} and the HI cloud catalog from the Leiden Dwingeloo Survey \citep[LDS;][]{hartmann97}, moving later to the HI Parkes All Sky Survey catalog \citep{putman02}. Many HVCs from these catalogs were studied unsuccessfully for optical emission \citep{davies02,simon02,Willman02,Hopp03,siegel05,hopp07}.  




The Galactic Arecibo $L$-Band Feed Array HI (GALFA-HI) and  Arecibo Legacy Fast Arecibo $L$-band Feed Array (ALFALFA) surveys
provide a new opportunity for discovering Local Volume galaxies within $\sim$10 Mpc.  
 Their spatial resolutions ($\sim$4$'$ for both) and sensitivity ($M_{HI}$$\sim$10$^5$--10$^6$ M$_{\odot}$ at 1 Mpc for both) are well-matched to the expected HI properties of star-poor Local Group low-mass halos \citep{Giovanelli05,Peek11}. Catalogs of ``ultra-compact" high velocity HI clouds, considered to be the best Local Volume galaxy candidates, have been published for both GALFA-HI's Data Release 1 \citep[58\% of the full survey,][]{Peek11} and ALFALFA's $\alpha$.40 HI Source Catalog  \citep[40\% of the full survey]{Haynes11}.  



\citet[][S12 hereafter]{Saul12} identified a set of 27 UCHVC ``galaxy candidates" from their GALFA-HI compact cloud catalog.  These objects were selected based on their velocity ($|V_{LSR}|$$>$90 km s$^{-1}$) and lack of association with any known HVC complex or galaxy. Most were barely resolved in the survey data, with a median size of 5'.  \citet[][A13 hereafter]{Adams13} extracted 59 UCHVCs from ALFALFA's $\alpha$.40 catalog based on their velocity ($|V_{LSR}|$$>$120 km s$^{-1}$) and size; the median size of the ALFALFA UCHVC catalog objects is 12', larger than the GALFA-HI candidates. Nonetheless, there 
is some overlap between the GALFA-HI and ALFALFA UCHVC compilations---eleven ALFALFA UCHVCs are in the GALFA-HI compact cloud catalog, but only one of these is in the ``galaxy candidate" group identified by the GALFA-HI team. These differences indicate the difficulty in defining what may be an intrinsically heterogeneous class of objects. Between the two catalogs, there are a total of 85 unique UCHVCs.




The spatial distribution and physical properties of these UCHVCs appear broadly consistent with the hypothesis that some of them are inhabited by optically-faint dwarf galaxies in the Local Volume \citep{giovanelli10, Adams13, faerman13}.  \citet{garrisonkimmel14} also predict that a subset of the A13 UCHVCs are Local Group inhabitants, and that the radial velocities of individual UCHVCs can serve as an initial guess as to which are most likely to host Local Group dwarf galaxies.

That one dwarf galaxy has already been discovered as an optical counterpart to an ALFALFA UCHVC supports the hypothesis that some UCHVCs inhabit Local Volume dark matter halos.  The gas-rich Leo P dwarf galaxy resides at D$\sim$1.7 Mpc, has a $M_{V}$=$-$9.4 and contains $M_{\rm HI} \sim$ 3.6 $\times$ 10$^5$ M$_{\odot}$ \citep{giovanelli13,Rhode13,mcquinn13}.  Spectroscopic studies have indicated that it is an extremely metal-deficient galaxy \citep{Skillman13}, with a rotation speed of 15 km/s, suggesting it lives in a dark matter halo similar to those surrounding faint dwarf galaxies in the Local Group \citep{Bernstein14}. Inspired by this discovery, several groups have recently pursued a combination of archival SDSS and targeted optical observations of S12 and A13's UCHVCs, and discovered dwarf galaxies associated with three of them \citep[][hereafter B14 and T15, respectively]{Bellazzini14,Tollerud15}.  However, no complete UV/optical archival search of S12's and A13's 85 UCHVCs has yet been conducted.

In this paper we present the results of a comprehensive search for optical counterparts to UCHVCs, using both archival and new data. 
We complement this archival search with new optical imaging and spectroscopic follow-up of all of our viable dwarf galaxy candidates.  In  Section~\ref{sec:search} we discuss our systematic search for dwarf galaxy counterparts.  We present optical spectroscopy of our dwarf galaxy counterparts in Section~\ref{sec:spectra}, in order to confirm their association with the UCHVC.  In Section~\ref{sec:discuss} we discuss the overall properties of our new dwarf galaxies.  We present our conclusions in Section~\ref{sec:conclude}.

When necessary, we adopt a Hubble constant of $H_{0}$ = 69.3 km s$^{-1}$ Mpc$^{-1}$ from WMAP9 \citep{Hinshaw13}

\section{A Search for UCHVC Dwarf Counterparts}\label{sec:search}

In this section we describe our search of publicly available UV/optical data to find UCHVC dwarf galaxy counterparts.  In addition to this archival data, we present some complementary deep imaging (Section~\ref{sec:imaging}) to both search for dwarfs to deeper limits and to confirm dwarf candidates in shallow archival imaging.  Our imaging-based search has uncovered six dwarf galaxy candidates, all of which we present spectra for in Section~\ref{sec:spectra}. Five are confirmed to have optical velocities consistent with the HI velocity, and are thus true dwarf galaxy counterparts to the UCHVC.

\subsection{A Visual Search for Dwarf Counterparts in Archival Data}\label{sec:archive_search}

We have searched all available optical and ultraviolet imaging archives for counterparts to the UCHVCs.  
In the optical, these include the Digitized Sky Survey (DSS), the Sloan Digital Sky Survey \citep[SDSS Data Release 10; ][]{SDSSDR10}, the Subaru SMOKA data archive \citep{SMOKA} and the Canada France Hawaii Telescope's Megacam archive \citep[utilizing the MegaPipe data products;][]{Gwyn08}. In the ultraviolet, we have made extensive use of the Galaxy Evolution Explorer \citep[GALEX;][]{GALEX} archive, which is especially good for uncovering young star forming regions, as is expected for any gas-rich dwarf galaxy the UCHVCs may harbor.  We also searched for {\it Swift} UltraViolet and Optical Telescope (UVOT) imaging coincident with our UCHVC candidates, but none were found.  

Our search is focused on diffuse counterparts to the UCHVC sources, rather than resolved stars.  As our guide, we used the visual appearance of Leo~P in the SDSS archive -- blue in color (due to recent star formation), blobby and diffuse, with possible O-stars or unresolved HII regions.  When multiple optical bands are available, color images were made, facilitating a search for blue counterparts.  We confined our visual search for these features to within the $\sim$1 arcmin positional uncertainty of the UCHVC's position (see A13 and S12), rather than searching the entire extent of the HI cloud, similar to what has been done in past searches for optical counterparts to HI clouds \citep[e.g.][]{Haynes11}.  This corresponds to $\sim$290 pc at 1 Mpc.  Despite this formal $\sim$1 arcmin search radius, diffuse dwarfs that were $\sim$2-3 arcmin offset from the HI position would have been recognized during the visual inspection of the images, but none were found.  All sources in optical bands with an apparent size $\gtrsim$5" were flagged, but most of these extended objects appeared to be background objects and were discarded, with clear spiral/elliptical morphologies, rather than the diffuse nature that we expected.  Diffuse galaxies with known velocities that did not agree ($\Delta$$v$$>$1000 km s$^{-1}$) with the UCHVC velocity were also discarded.  Any available UV imaging was used to confirm and support dwarf galaxy candidate selection, as the presence of UV flux is expected from a young star forming galaxy.  Multiple members of our team searched each image by eye (DJS, DC, PB), with no disagreement between the searchers on dwarf candidates.

Table~\ref{table:archive} details the results of this archival search, which was supplemented by a handful of new images presented in Section~\ref{sec:imaging}.  For data collected from archives other than SDSS and DSS, we have noted the image exposure lengths (or typical survey exposure times) of our searched fields.  For instance, in the footnotes to Table~\ref{table:archive}, we note which UCHVC fields have coverage in the GALEX All-Sky Imaging Survey (AIS) or the Medium Imaging Survey (MIS); see \citet{Morrissey07} for details on these GALEX surveys.  The exposure times for the CFHT imaging are also presented in the footnotes of the table.

Our search uncovered six strong dwarf galaxy candidates, shown in Figures~\ref{fig:imgs} and \ref{fig:imgs2}, five of which we were able to confirm with optical spectroscopy were at the same velocity as the UCHVC cloud (see Section~\ref{sec:spectra}).  For these five dwarfs,  we assign names according to the discovering HI survey, in order of increasing right ascension; e.g. GALFA-Dw1, GALFA-Dw2, etc.  Two of these, GALFA-Dw1 and GALFA-Dw2, were identified previously by T15 as Pisces~A and Pisces~B, respectively.  A third, ALFALFA-Dw1, was identified and spectroscopically confirmed by \citet{Bellazzini15} and named SECCO 1 -- we confirm its optical velocity is consistent with the UCHVC velocity in Section~\ref{sec:spectra}.  Given that these dwarfs are not likely to be Local Group members (Section~\ref{sec:distance}) and future workers are likely to find large numbers of Local Volume dwarf galaxies \citep[e.g.][]{James14}, we prefer this nomenclature to, for instance, naming the systems after the corresponding constellation.  The current naming scheme also gives credit to the appropriate HI survey.

In the Appendix, we also note several clarifications uncovered during the archival search; including sources that have been previously identified in the literature, or that are associated with known systems.  We hope that these clarifications will help future, deeper searches for UCHVC counterparts in prioritizing fields.

\subsection{Supplementary Imaging}\label{sec:imaging}

In addition to our archival search for counterparts to the UCHVCs, we have also acquired some supplementary imaging of select systems, either to search for dwarf counterparts to deeper limits or to confirm tentative candidates in shallower archival images.  A log of these observations is presented in Table~\ref{table:imaging}, in the order that the data was taken.  

\begin{figure*}
\begin{center}
\mbox{ \epsfysize=4.3cm \epsfbox{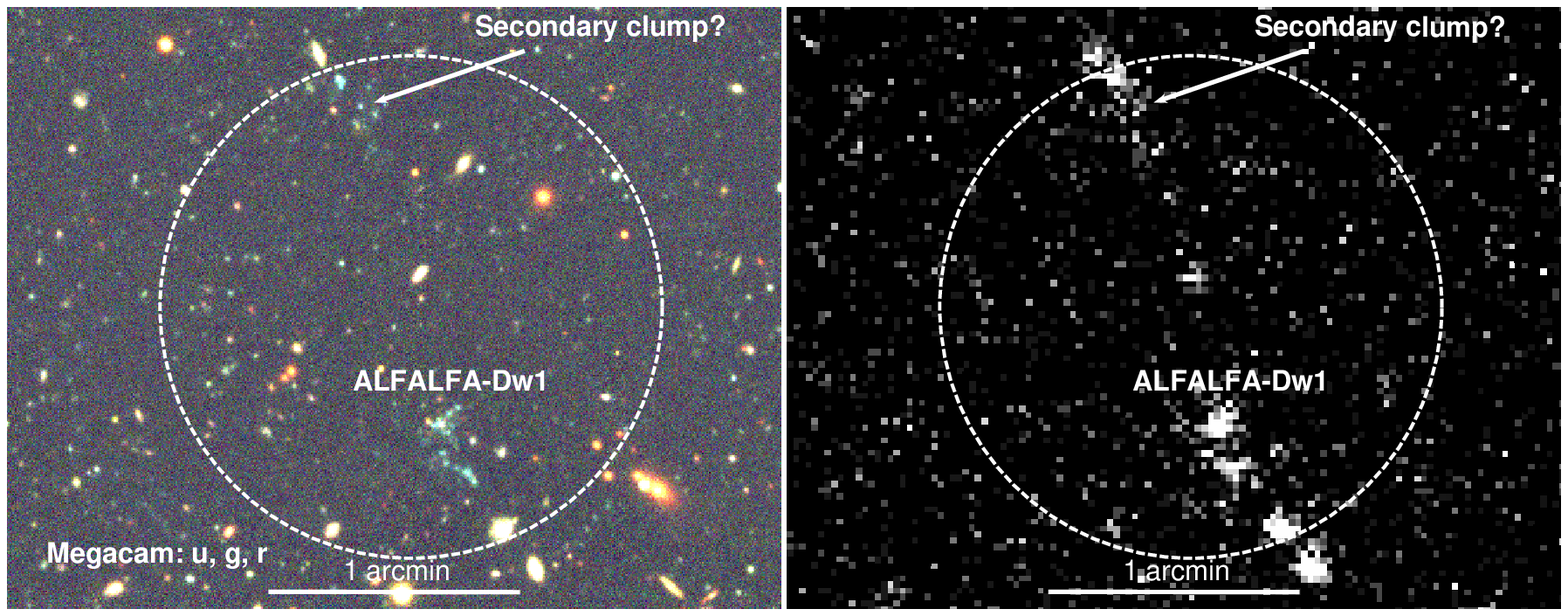}} 
\mbox{ \epsfysize=4.3cm \epsfbox{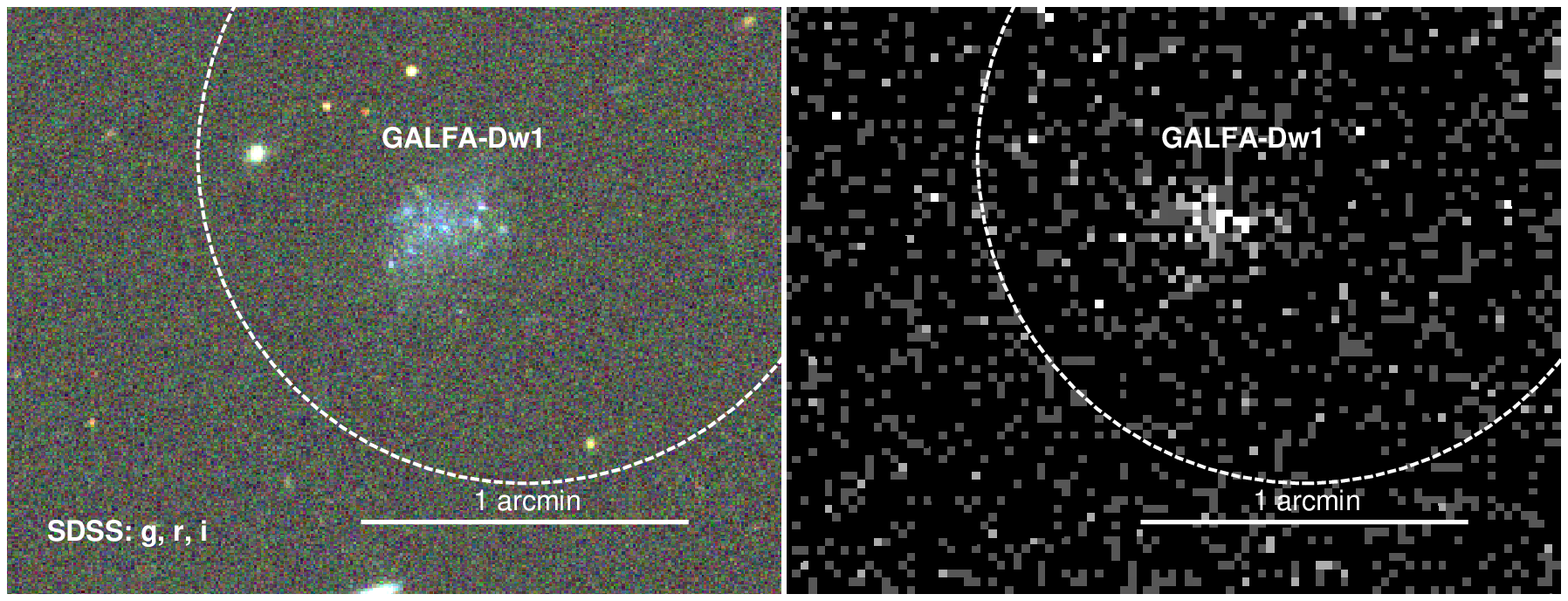}} 
\mbox{ \epsfysize=4.3cm \epsfbox{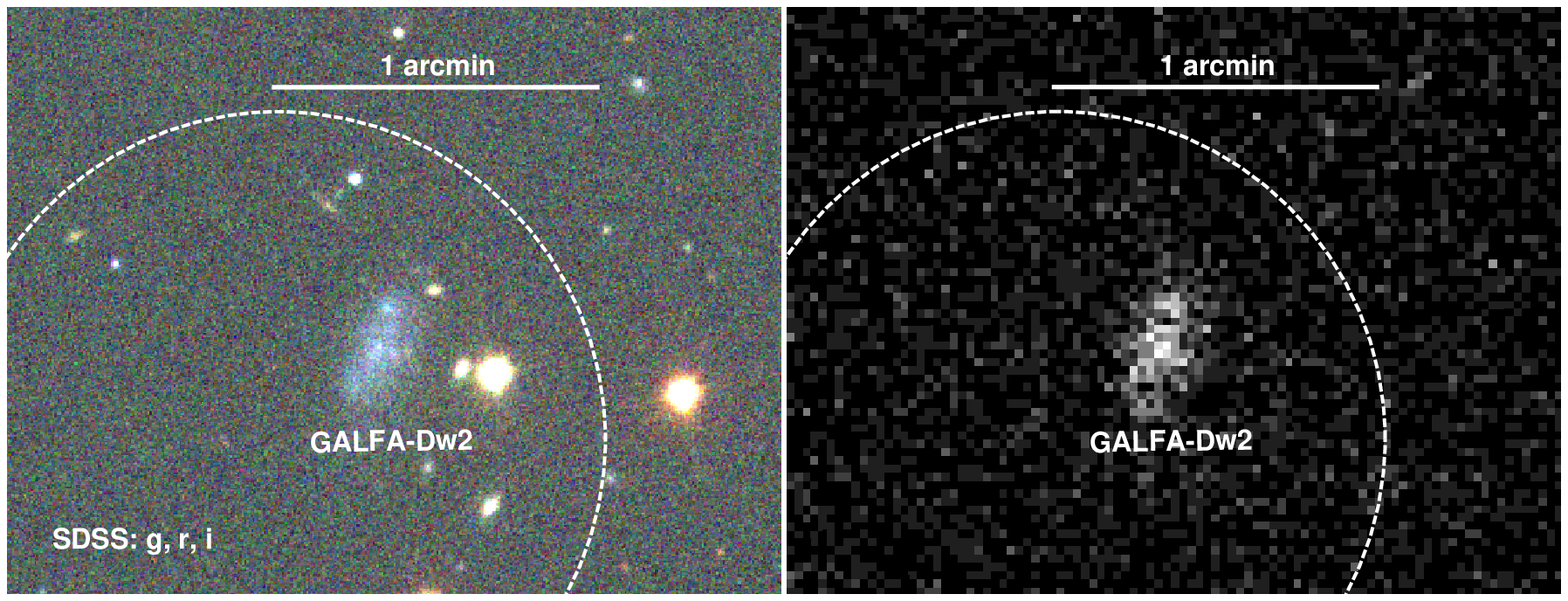}} 
\mbox{ \epsfysize=4.3cm \epsfbox{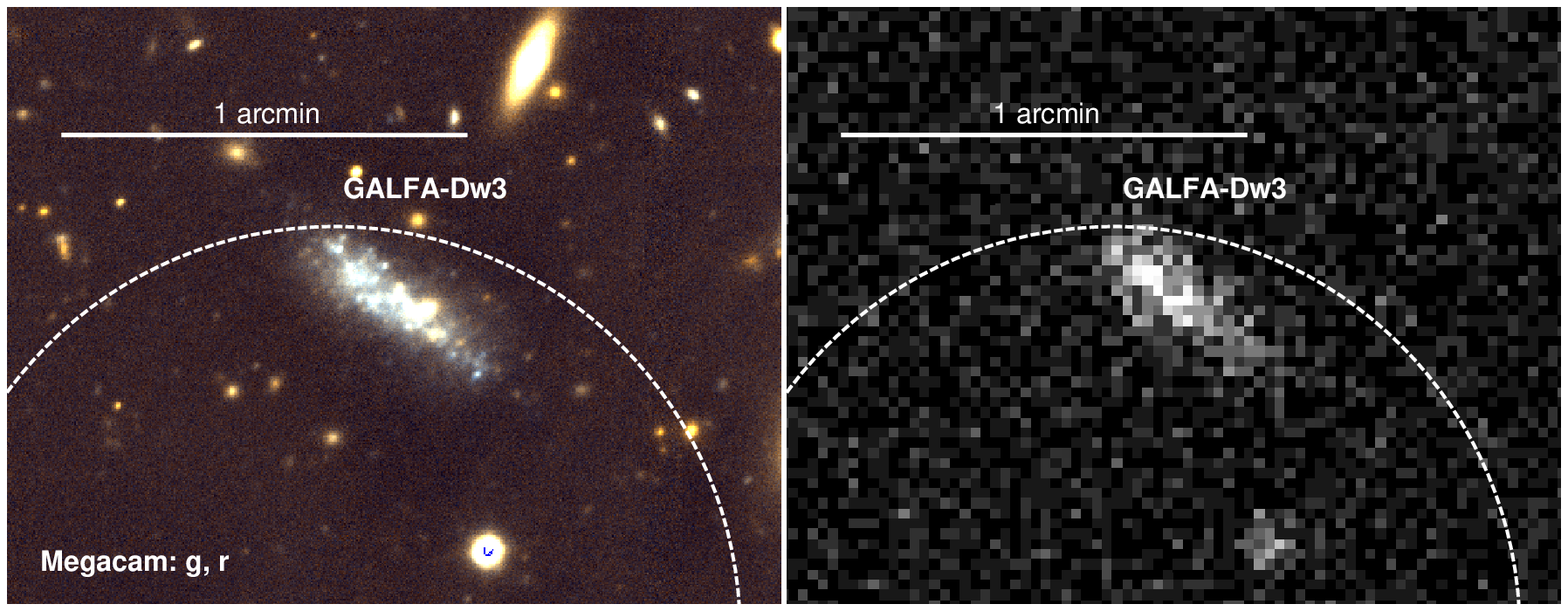}} 
\mbox{ \epsfysize=4.3cm \epsfbox{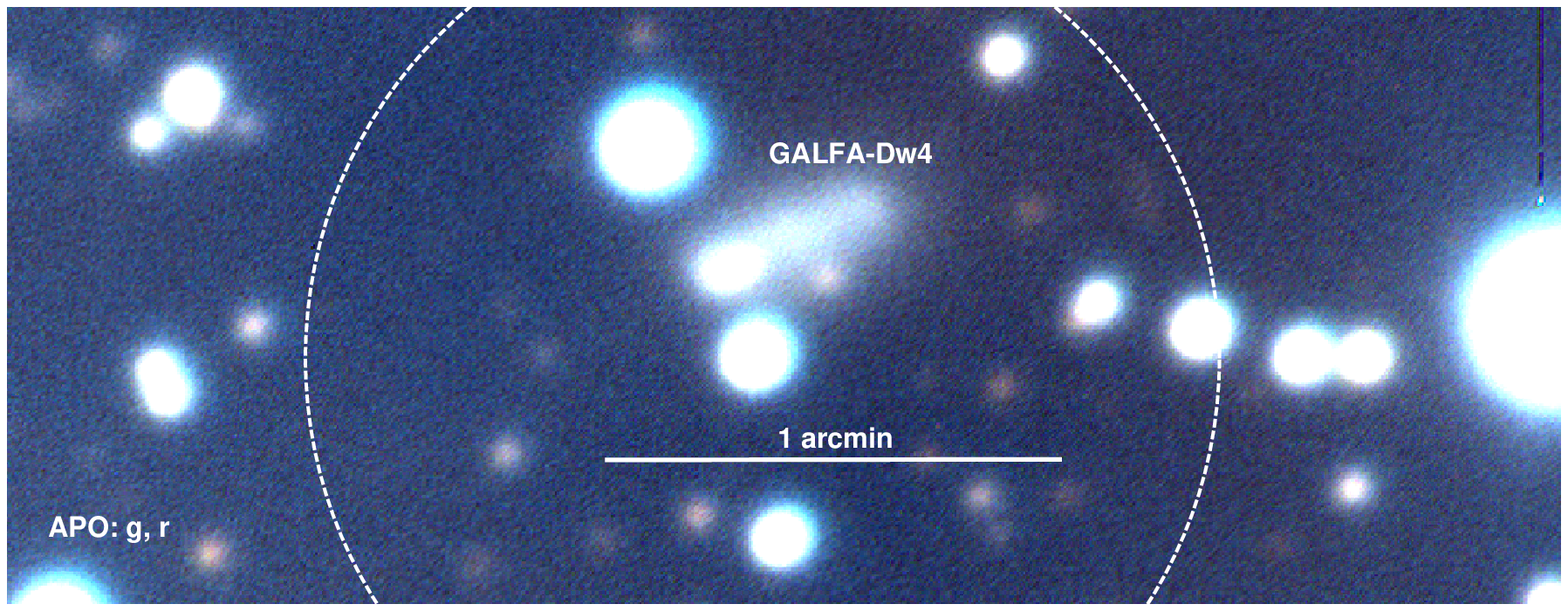}} 
\caption{  Color optical images, as well as GALEX NUV data (right panel, when available) of the five dwarf galaxies uncovered in archival data with optical velocities consistent with their coincident UCHVC cloud.  GALFA-Dw4 is $\sim$10 degrees from the plane of the Milky Way, and is behind $\sim$1.5--2 magnitudes of extinction; there is no GALEX UV data.   The three-color image for GALFA-Dw4 was constructed using the two available bands, and averaging their flux to create the green channel of the image.   All of the dwarf galaxies appear blue in the optical, and have strong GALEX detections, indicating a young stellar population.  The dashed circular region roughly represents the HI positional uncertainty.  North is up and East is left.   \label{fig:imgs}}
\end{center}
\end{figure*}

{\it HVC274.68+74.70$-$123 (ALFALFA-Dw1) -- }  We have collected complementary information on this strong dwarf candidate, which we have dubbed ALFALFA-Dw1 (see Figure~\ref{fig:imgs}). Images were taken on 11 June 2013 (UT) utilizing Magellan/Megacam \citep{McLeod15} with exposures of 8$\times$300 sec in $g$ and 7$\times$300 in $r$; the conditions were non-photometric. This object was also flagged by B14 as a possible dwarf counterpart to the UCHVC, and was confirmed spectroscopically in \citet{Bellazzini15}.  As the CMD of this dwarf has been presented by B14, we do not reproduce it here, but we draw very similar conclusions. Beyond the information reported on this source by \citet{Bellazzini15}, we note a possible second knot of blue optical sources (and UV emission) roughly $\sim$2' Northeast of the main body of this dwarf (see Figure~\ref{fig:imgs}).  While we do not address this secondary knot further in this work, our upcoming Hubble Space Telescope observations will have full coverage of the ALFALFA-Dw1 field (PID: 13735; PI: Sand).  Based on the lack of a tip of the red giant branch (TRGB), we can tentatively place this dwarf beyond $D$$\sim$3 Mpc.  We also present our optical velocity measurement in Section~\ref{sec:spectra}, and discuss ALFALFA-Dw1's physical properties in later sections.

{\it HVC351.17+58.56+214/GALFA 215.9+04.6+205 -- }  This UCHVC was flagged by both teams (S12, A13), and was mentioned by A13 in particular for having HI properties similar to Leo~T and Leo~P, with both a high column density ($N_{HI}>10^{19}$ cm$^{-2}$) and a small angular diameter ($<$7').  Archival observations include SDSS and GALEX AIS NUV, FUV imaging.  We imaged HVC351.17+58.56+214 on 27 April 2014 (UT) with Magellan/Megacam.  The total exposure time was 6$\times$300 s in both the $g$ and $r$ band, taken in photometric conditions, with a limiting magnitude of $r_{0}$$\sim$25.9 mag \citep[see][for data reduction details]{Crnojevic14}.  No clustering of red or blue stars is apparent in the color magnitude diagram at the position of the HI cloud, and we estimate a lower limit of $D$$\sim$4 Mpc for any coincident dwarf galaxy.

{\it GALFA 044.7+13.6+528 (GALFA-Dw3) -- }Inspection of the DSS image at the position of this HI cloud reveals a clear low surface brightness dwarf candidate, which we have dubbed GALFA-Dw3.  As such, we imaged the field on 26 October 2014 (UT) with Magellan/Megacam with exposures of 8$\times$300s in $r$ and 7$\times$300s in $g$.  The night was not photometric, and without SDSS coverage, we used the AAVSO Photometric All-Sky Survey \citep[APASS;][]{Henden12} to roughly calibrate the field, with a limiting magnitude of $r_{0}\sim$25.5 mag.  The dwarf is only semi-resolved and crowded in the central regions, and we are unable to accurately determine the stellar population, or measure a TRGB distance. Based on this and the limiting depth of the observations, we conservatively estimate that this dwarf can not be closer than $D$$\sim$3 Mpc.  The optical velocity of this dwarf is presented in Section~\ref{sec:spectra}.

{\it HVC131.90$-$46.50$-$276 --}  This UCHVC was imaged on 17 November 2014 (UT) with the Seaver Prototype Imaging camera (SPICAM) at the 3.5 m Astrophysical Research Consortium Telescope at Apache Point Observatory (APO).  Images in $g$ and $r$ band were taken with exposure times of 2$\times$540s in each band.  The data were reduced in a standard way and data from each band were combined and astrometrically corrected using {\sc SWarp}\footnote{http://astromatic.iap.fr/software/swarp/} and {\it astrometry.net} \citep{Lang10}. The data were photometrically calibrated directly to the SDSS system with overlapping stars in the field.  Due to poor seeing ($\sim$4\farcs5), the data have a depth of $g$$\sim$23.3 and $r$$\sim$23.5 mag, with no sign of a diffuse counterpart to the UCHVC, or of an over density of point sources.

{\it GALFA 086.4+10.8+611 (GALFA-Dw4) -- } DSS imaging at the position of this UCHVC showed a clear candidate dwarf galaxy.  Follow up imaging of this dwarf candidate were taken with APO/SPICAM on 17 November 2014 (UT) in $g$ and $r$ band, with exposures of 4$\times$450s in each filter.  The data were reduced identically to that of HVC131.90$-$46.50$-$276 described above.  The data were photometrically calibrated using zero points and color terms derived from two standard star fields taken during the night at different airmasses.  We again note that the seeing was poor ($\sim$4") and variable, and so no resolved structure is apparent in this dwarf beyond two `blobs'.  A spectrum and optical velocity is presented in Section~\ref{sec:spectra}.

\section{Spectroscopy}\label{sec:spectra}

\begin{figure*}
\begin{center}
\mbox{ \epsfysize=4.3cm \epsfbox{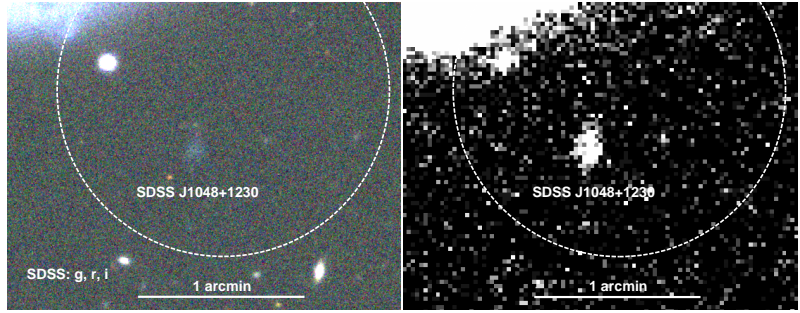}} 
\caption{  An SDSS 3-color image of the dwarf galaxy candidate associated with GALFA 162.1+12.5+434, along with the corresponding GALEX NUV data.  The dashed circular region roughly represents the HI positional uncertainty.  Follow up spectroscopy showed a single, weak emission line.  If this line is associated with H$\alpha$, the galaxy would have a velocity of  $\sim$5700 km s$^{-1}$, and is likely not related to GALFA 162.1+12.5+434.  It is possible that the UCHVC GALFA 162.1+12.5+434 is associated with the nearby galaxy NGC~3389 (just visible in the Northeast portion of the image), with a velocity difference of $\sim$800 km s$^{-1}$.  \label{fig:imgs2}}
\end{center}
\end{figure*}

\begin{figure*}
\begin{center}
\mbox{ \epsfysize=6.0cm \epsfbox{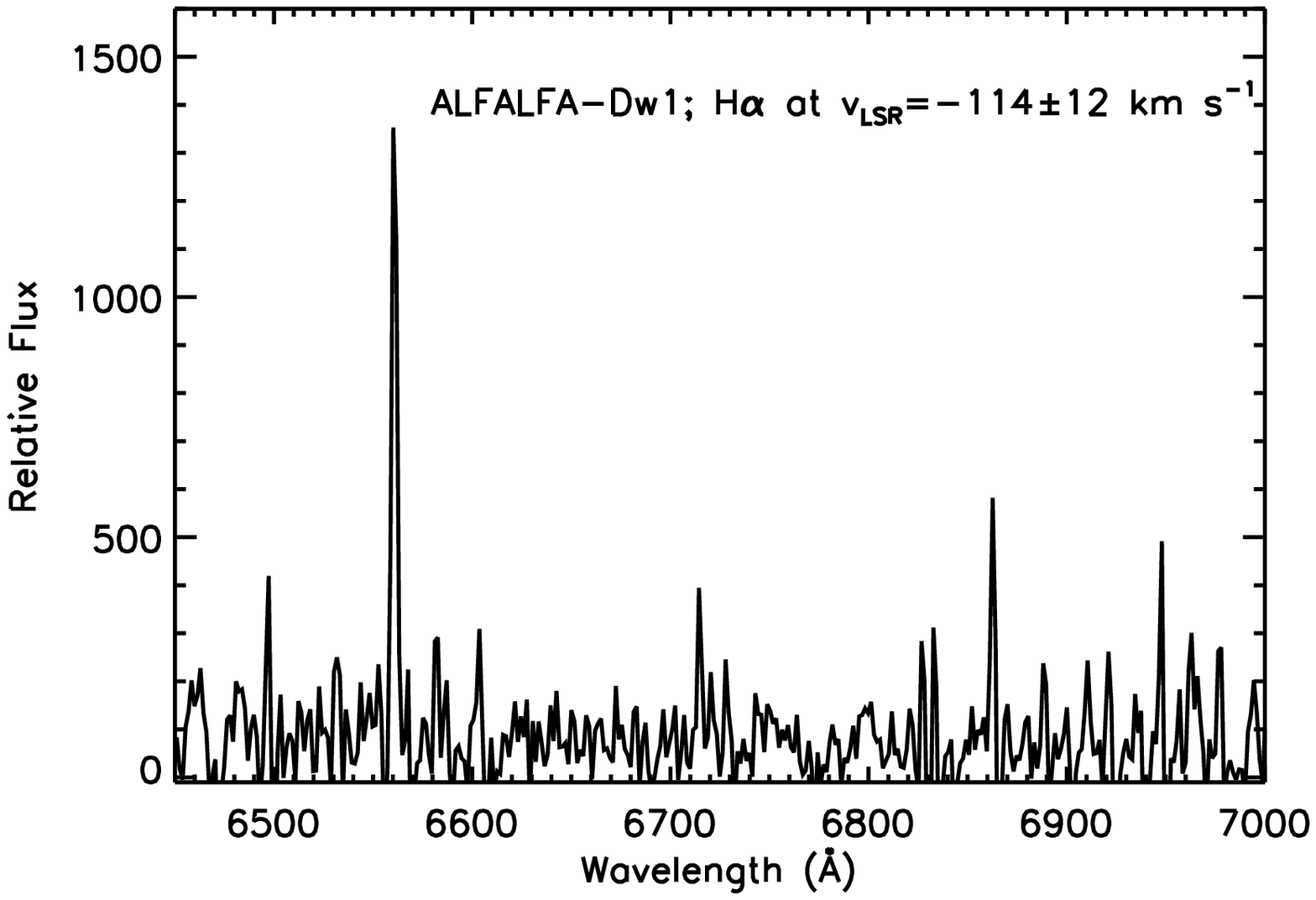}} 
\mbox{ \epsfysize=6.0cm \epsfbox{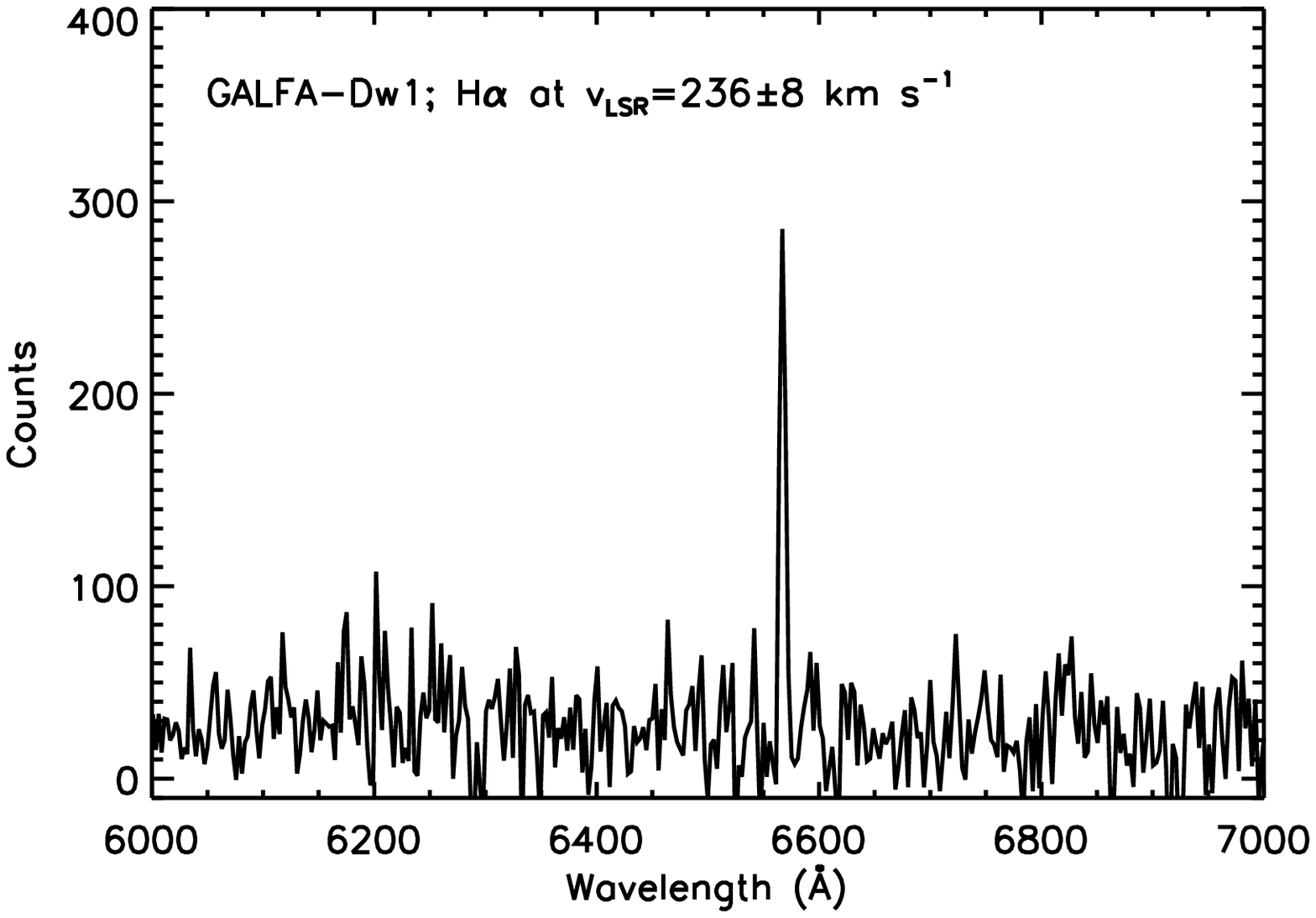}} 
\mbox{ \epsfysize=6.0cm \epsfbox{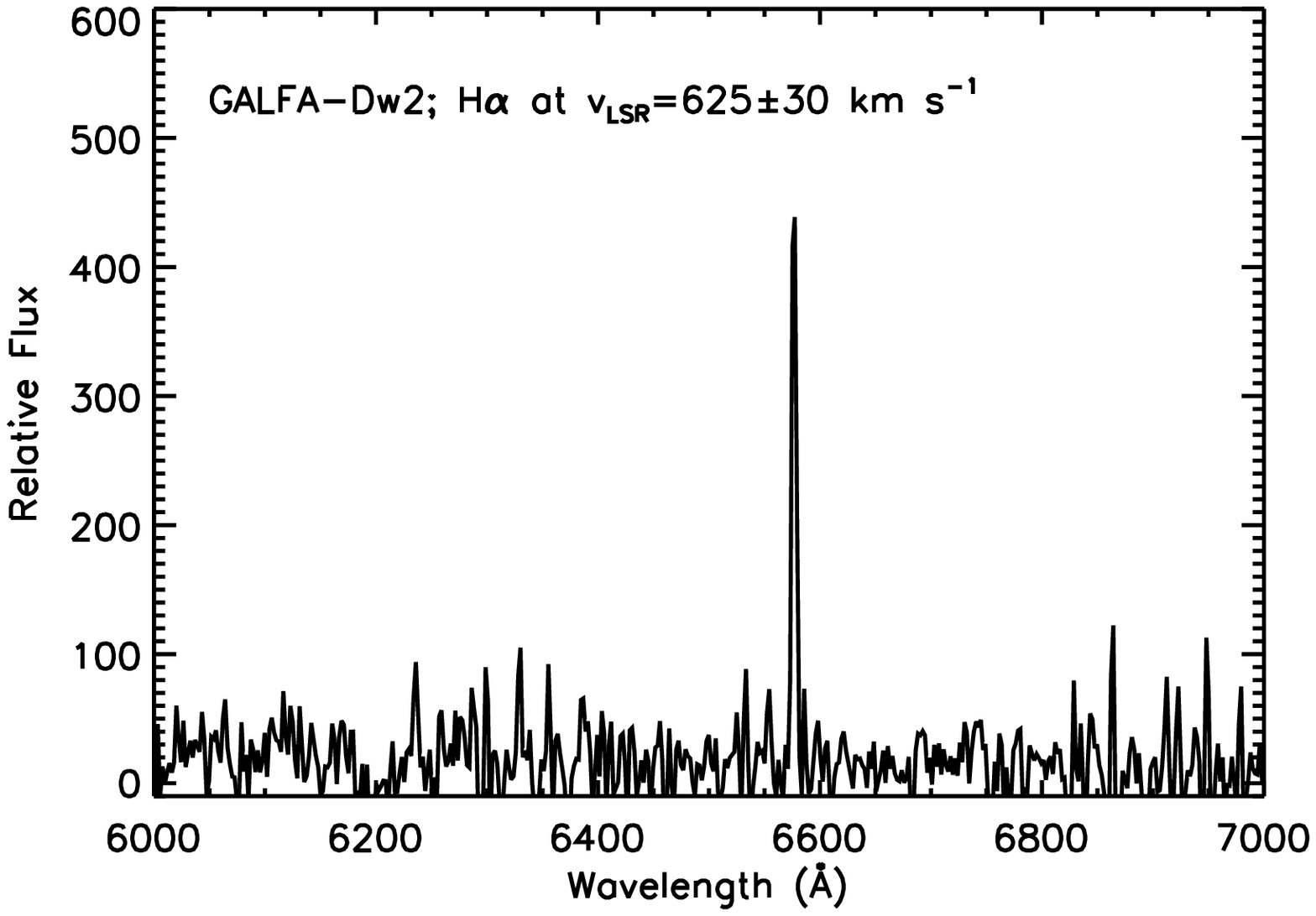}} 
\mbox{ \epsfysize=6.0cm \epsfbox{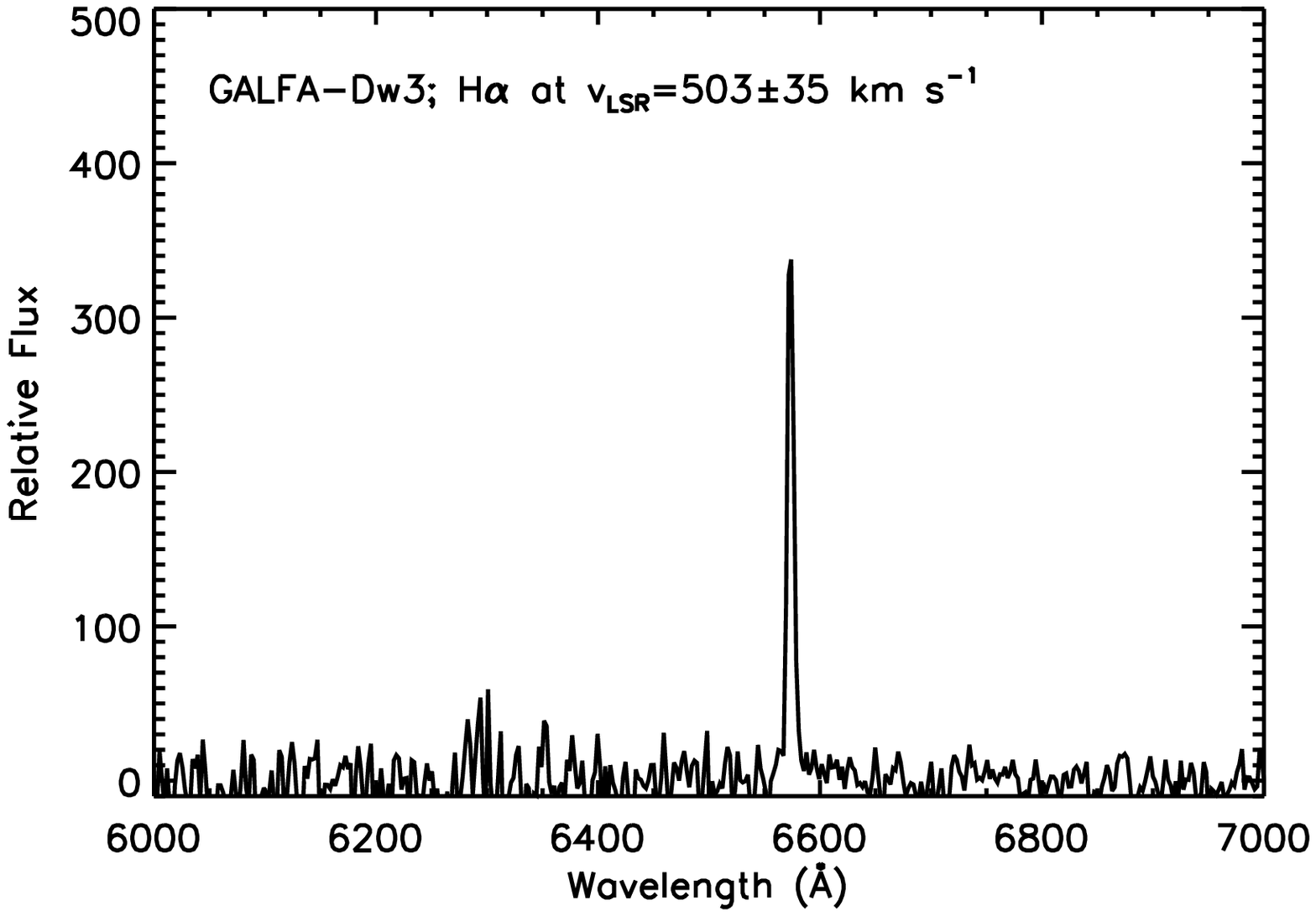}} 
\mbox{ \epsfysize=6.0cm \epsfbox{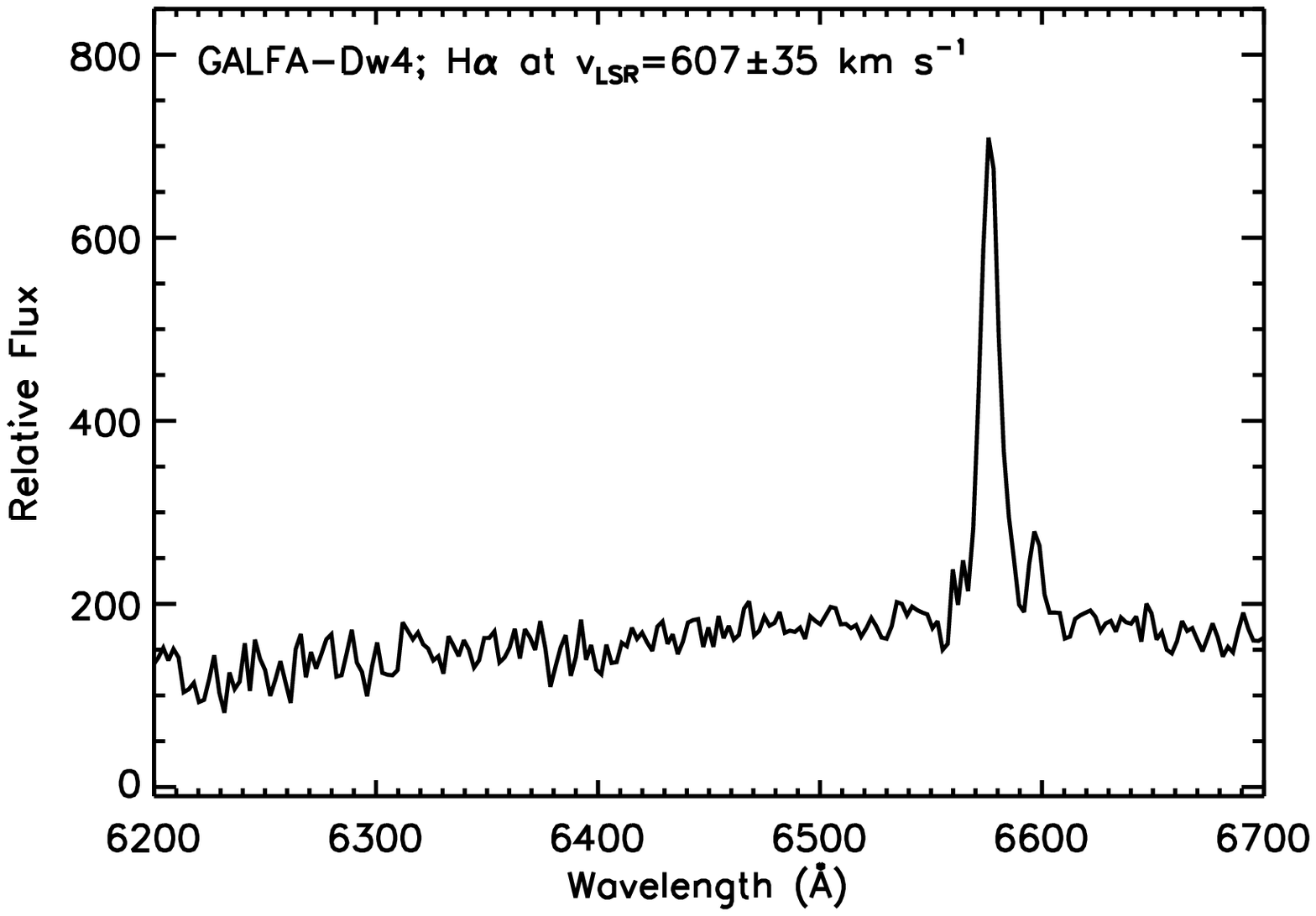}} 
\caption{ Spectroscopic cutouts of the five UCHVC counterparts with H$\alpha$ velocities consistent with that of HI.  Further details for each observation, and other detected optical lines, can be seen in Table~\ref{table:spec} and Section~\ref{sec:spectra}.  
\label{fig:specs}}
\end{center}
\end{figure*}

We have obtained spectroscopic observations for all six of the strong dwarf galaxy candidates identified in our search.  Of these six (see Figures~\ref{fig:imgs} and \ref{fig:imgs2}), five clearly exhibit H$\alpha$ with a velocity consistent with that of the UCHVC, confirming them as dwarf counterparts.  The sixth object (GALFA 162.1+12.5+434), with a single faint emission line, is likely not associated with the coincident UCHVC.  In this section, we detail our spectroscopic observations of all of our dwarf galaxy candidates.  An observational log can be see in Table~\ref{table:spec}, and the spectra of the five confirmed dwarfs are presented in Figure~\ref{fig:specs}.

{\it ALFALFA-Dw1 (HVC274.68+74.70$-$123)} was observed with the Inamori-Magellan Areal Camera and Spectrograph \citep[IMACS;][]{IMACS} on the Magellan Baade telescope on 17 June 2014 (UT).  We obtained five 1800s exposures with the 300 line grating and a 0\farcs7 slit, with wavelength coverage from $\sim 3300$--9600 \AA . A resolution of $\sim$2.7\AA~was measured from unblended sky lines near the wavelength of $H\alpha$.
The slit was oriented along the dwarf, revealing spatially extended H$\alpha$ emission across $\sim$12", along with H$\beta$ and the [O III] $\lambda \lambda$4959, 5007 doublet.  An extraction of the full emission yielded an H$\alpha$ radial velocity (LSR) of $-$114$\pm$ 12 km s$^{-1}$, fully consistent with the HI velocity of $-$123 km s$^{-1}$.  We clearly identify the two H$\alpha$ knots discussed by \citet{Bellazzini15}, but do not recover the large velocity difference that they see, and which they attribute to mis-centering in the slit.


{\it GALFA-Dw1 (GALFA source 003.7+10.8+236)}, also known as Pisces~A was observed on 29 June 2014 (UT) using the Goodman High-Throughput Spectrograph \citep{Clemens04} on the SOAR 4.1-m telescope. We obtained five 900 sec exposures with the 400 l mm$^{-1}$ grating and a 1.03\arcsec slit, yielding a resolution of 5.7\AA\ and wavelength coverage from $\sim 3100$--7000 \AA. The longslit was oriented at the parallactic angle at the start of the first exposure, and covered both the brightest knot and the main body of the candidate galaxy. Spatially extended emission consistent with the expected
position of H$\alpha$ was apparent on the resulting two-dimensional spectra over the location corresponding to the galaxy. No spectral features were associated with the knot itself. Guided by the distribution of the putative H$\alpha$ emission, we extracted spectra in a 4.8\arcsec~aperture centered
on the spatial H$\alpha$ peak emission. 
The only significant emission line in the spectra is the bright H$\alpha$ line, from which we derive a radial velocity (LSR) of 236$\pm$8 km s$^{-1}$ through cross-correlation with a rest-frame emission template.  This velocity is consistent with both the GALFA HI velocity (235.37 km s$^{-1}$) and the optical H$\alpha$ velocity reported by T15.

{\it GALFA-Dw2 (GALFA source 019.8+11.1+617)}, also known as Pisces B, was observed on 21 October 2014 (UT) with the Blue Channel Spectrograph on the MMT \citep{BCS}.  A single 900 sec exposure was obtained with the 300 l mm$^{-1}$ grating and a 1\arcsec~slit.  The central wavelength was 5300 \AA, with an order-blocking filter ($<$3600 \AA), yielding a wavelength range of 3600--8000 \AA.  A resolution of $\sim$7\AA~was measured from unblended sky lines.
Spatially extended emission was detected across the face of the dwarf, with emission lines visible in two extended knots.  Emission lines corresponding to H$\alpha$, H$\beta$ and the [O III] $\lambda \lambda$4959, 5007 doublet are visible in the spectrum.  We extracted a spectrum from one of the knots, and measure a H$\alpha$  radial velocity of $v_{LSR}$=625$\pm$30 km s$^{-1}$.  Again, this velocity is consistent with both the GALFA HI velocity (611.63 km s$^{-1}$) and the optical H$\alpha$ velocity reported by T15.

{\it GALFA-Dw3 (GALFA source 044.7+13.6+528) and GALFA-Dw4 (GALFA source 086.4+10.8+611)} were observed with the 3.5 m Astrophysical Research Consortium Telescope at Apache Point Observatory (APO) on 18 November 2014 (UT) using the Dual Imaging Spectrograph (DIS).  The APO+DIS observations were taken with a 1\farcs5 slit and the B400/R300 gratings on the red/blue arms of the instrument; wavelength calibration and flat fields were taken at the science position.  The position angle for each observation was oriented to maximize the amount of the galaxy (and any knots it may contain) in the slit.  
The spectrum of GALFA-Dw3 exhibits a clear H$\alpha$ line, along with faint H$\beta$ and [O III] $\lambda \lambda$4959, 5007. With our spectroscopic setup, the H$\alpha$ is concentrated in a single knot, and does not extend across the body of the dwarf; we measure an H$\alpha$ velocity of $v_{LSR}$=503$\pm$35 km s$^{-1}$, consistent with the UCHVC ($v_{LSR}$=528.59 km s$^{-1}$).
The spectrum of GALFA-Dw4 exhibits H$\alpha$ emission from two distinct and spatially extended knots; also visible are H$\beta$, [O III] $\lambda \lambda$4959, 5007, higher order hydrogen Balmer lines, [N II] $\lambda$6584 and several He I lines.  The H$\alpha$ velocity, $v_{LSR}$=607$\pm$35 km s$^{-1}$, is consistent with the UCHVC at $v_{LSR}$=614.53 km s$^{-1}$.

The faint, blue optical source coincident with {\it GALFA 162.1+12.5+434} (see Figure~\ref{fig:imgs2}) was observed on 24 December 2014 (UT) with the Blue Channel Spectrograph on the MMT, again with the 300 l mm$^{-1}$ grating and a 1\arcsec~slit, with a total exposure time of 2$\times$900 s.  Spatially extended continuum emission was detected across the face of the target, with a single faint emission line.  If this emission line is H$\alpha$, then the velocity of the source is 5700$\pm$200 km s$^{-1}$, where we have taken the mean velocity of the two exposures and their spread as the central value and uncertainty, respectively.  If this is the correct line identification, then the blue source seen in Figure~\ref{fig:imgs2} is not associated with the HI source GALFA 162.1+12.5+434 ($v_{LSR}$=435.67 km s$^{-1}$), and we leave it labeled by the SDSS name of the blue source -- SDSS J1048+1230 -- in Table~\ref{table:spec}. We will not discuss it further in this work.

\section{Properties of the Dwarfs}\label{sec:discuss}

\subsection{Distance Estimates}\label{sec:distance}

The distances to the UCHVC dwarfs are unknown, and will likely require Hubble Space Telescope depth and image quality to determine TRGB distances.  Nonetheless, we can estimate upper and lower limits on their distances in order to infer their distance-dependent physical properties in the rest of this section.  Our assumed distance range for each dwarf is listed in Table~\ref{table:properties}.

Both GALFA-Dw1 and GALFA-Dw2 were studied by T15, and we adopt their lower distance bound for our estimates.  For GALFA-Dw1, T15 assumed that the lower bound on the distance was D=1.7 Mpc, based on a comparison of point sources in GALFA-Dw1 with respect to Leo~P's main sequence stars.  For GALFA-Dw2, T15 assumed a lower distance limit of D=3.5 Mpc based on their deep ground-based imaging.  As an upper limit for the distance to these dwarfs we adopt a distance which is a factor of 1.5 above the Hubble flow value for each dwarf, in order to be conservative -- D=5.3 Mpc for GALFA-Dw1 and D=13.4 Mpc for GALFA-Dw2.  

Similarly, ALFALFA-Dw1 was recently studied by \citet{Bellazzini15}, and we broadly adopt their distance estimate range.  Based off of their Large Binocular Telescope color magnitude diagram (and the lack of a TRGB), they estimate a lower distance limit of D=3 Mpc.  Our own Magellan/Megacam color magnitude diagram agrees with this estimate.  However, ALFALFA-Dw1 is projected onto the Virgo cluster, and it is possibly a faint, star forming member; this would imply a distance of up to D=18 Mpc \citep[using a distance to the Virgo Cluster of D=16.5 Mpc, with an intrinsic spread in line of sight distances of $\sim$1.5 Mpc;][]{Mei07}.

For both GALFA-Dw3 and GALFA-Dw4, we take a simple approach, assuming that their distance lies somewhere between D=3 Mpc (based off of the Magellan/Megacam image depth for GALFA-Dw3) and a factor of 1.5 beyond their Hubble flow distance -- D=11.4 Mpc for GALFA-Dw3 and D=13.3 Mpc for GALFA-Dw4.  

\subsection{Luminosity and Half-Light Radius}\label{sec:l_and_size}

We estimate the magnitude and luminosity of our new dwarfs directly from elliptical aperture photometry, similar to that presented in \citet{Sand14}.  We choose an elliptical aperture for each dwarf based on all available imaging (in different bands) such that no dwarf flux extends beyond the aperture.  Note that the position angle and ellipticity of the aperture is not constrained directly, but is simply chosen to be in the visible direction of the light profile.  Care is taken to avoid bright nearby stars within the aperture, although some contamination is inevitable.  The aperture position is adjusted by centroiding on the $r$-band image within an initial elliptical aperture, placed at the approximate center of the dwarf.  We use this optical centroid position as the position of the dwarf, which we report in Table~\ref{table:properties}, along with the other parameters derived in this Section.  

For our optical data, we randomly place $\sim$30--50 apertures of equal area to that used to measure the dwarf flux throughout the field, in order to estimate the background.  This accounts for interloping stars/galaxies in the dwarf aperture in a statistical sense.  We report our measured $r_{0}$ and $(g-r)_{0}$ color for each dwarf in Table~\ref{table:properties}, after accounting for Galactic extinction \citep{Schlafly11}.  For our GALEX UV data, we use the same elliptical aperture to measure the dwarf flux as in the optical, but instead use an elliptical annulus $\sim$20-30" beyond the radius of the dwarf to measure the background.  We chose this methodology for the UV data because of the strongly varying image quality across the GALEX focal plane.

The color of the dwarf galaxies are all $(g-r)_{0}$$\lesssim$0.3 mag.  Comparison with a set of single stellar population colors as a function of age \citep{Bressan12} suggests this corresponds to stellar ages of $\lesssim$1 Gyr. For those dwarfs with a color of $(g-r)_{0}$$\sim$0 mag (e.g. ALFALFA-Dw1, GALFA-Dw4, and even the others within the uncertainties) the single stellar population age is nearer $\sim$100 Myr.  Higher precision photometry is necessary to further constrain the stellar population of these dwarfs.

The half light radius of each dwarf was measured from the $r$-band data.  The photometric aperture of each dwarf (and the background apertures) was closed down until half of the total dwarf flux was enclosed, taking into account background uncertainties and the uncertainty in the total magnitude of each dwarf.

We can compare our results for GALFA-Dw1 and GALFA-Dw2 directly with T15, keeping in mind that the current work utilizes SDSS imaging,  while T15 used deeper ground-based data.  Both sets of measurements are consistent for GALFA-Dw1, to within the 1-$\sigma$ uncertainties, while the measurements for GALFA-Dw2 are consistent within 2-$\sigma$.  Given the diffuse nature of these dwarfs, and the different quality datasets, we consider this level of agreement satisfactory.  Similarly, comparing our measurements with those of ALFALFA-Dw1 presented in B15, we find agreement within the 1-$\sigma$ uncertainties.

\begin{figure*}
\begin{center}
\mbox{ \epsfysize=6.0cm \epsfbox{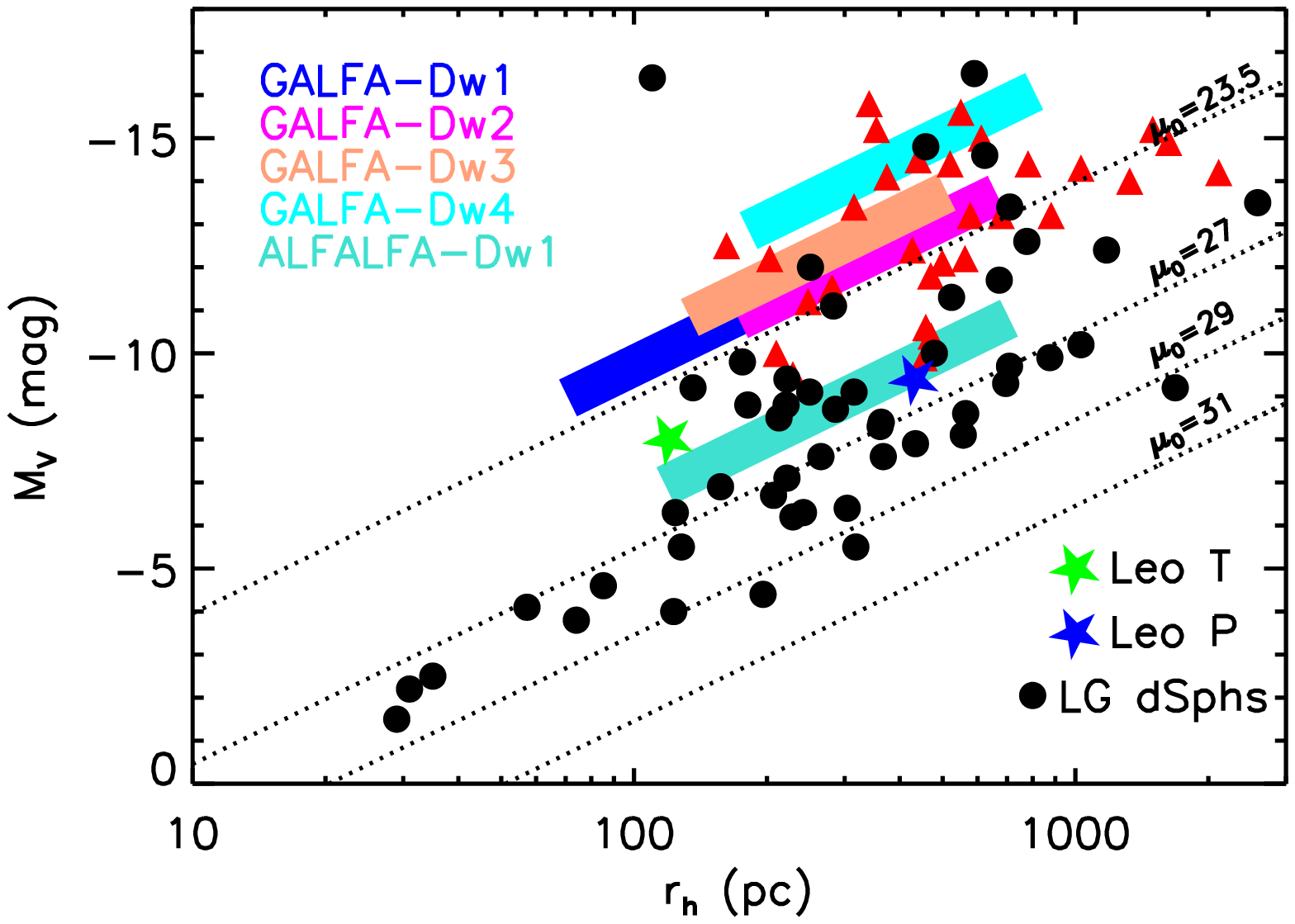}} 
\mbox{ \epsfysize=6.0cm \epsfbox{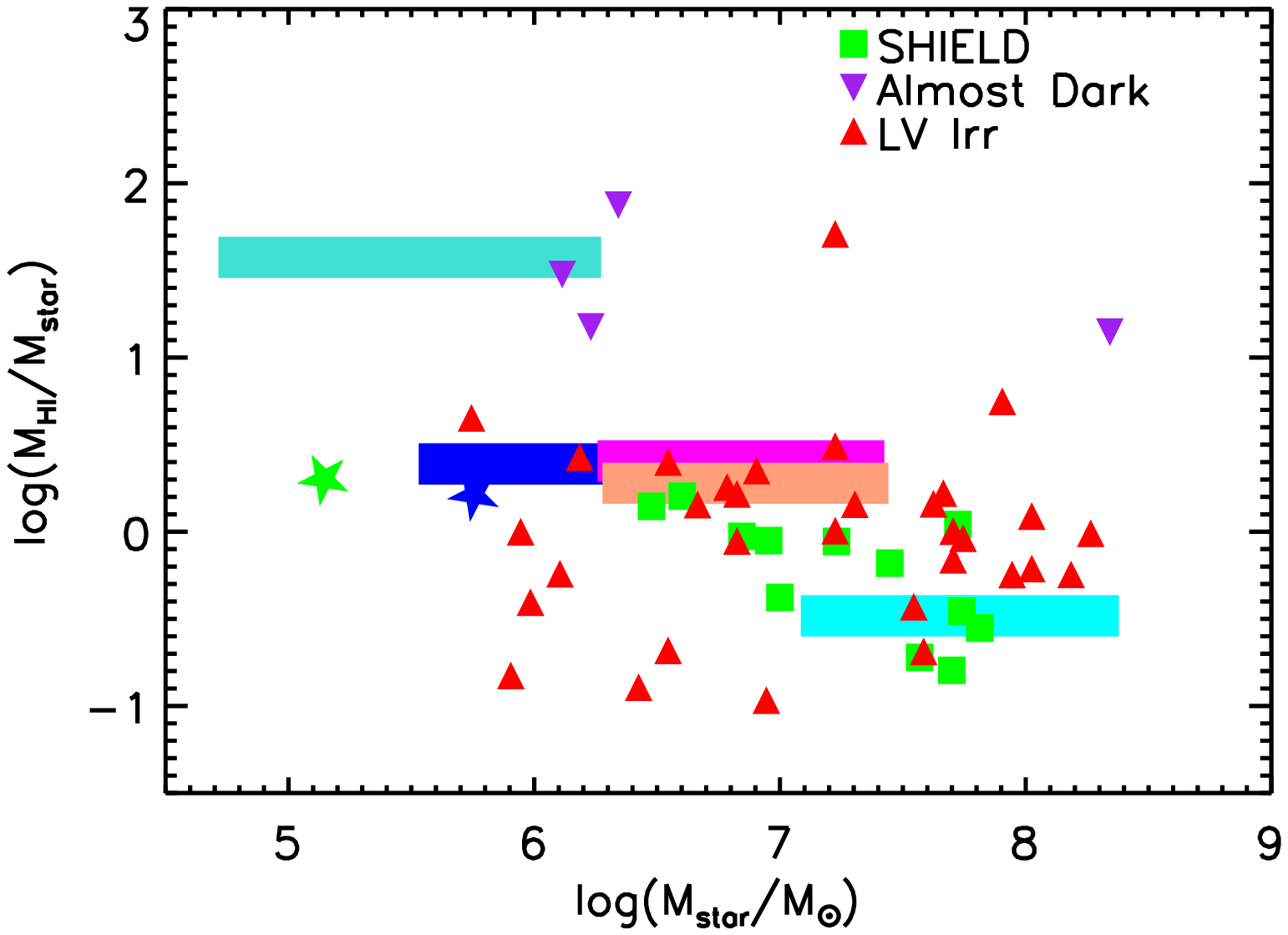}} 
\caption{  Left -- Absolute magnitude as a function of half light radius for both Local Group dwarf spheroidals (around the Milky Way and M31) and dwarf irregular galaxies in the Local Volume.  All data were taken from \citet{McConnachie12} and \citet{Sand12}, except for that of Leo~P, which is from \citet{mcquinn13}.  We draw lines of constant central surface brightness (assuming an exponential profile) in order to emphasize that there are still likely to be selection effects associated with all of the known dwarf galaxy populations, including the current search -- see discussion in Section~\ref{sec:l_and_size}.  The UCHVC dwarfs uncovered by our search, marked by colored bands corresponding to their distance uncertainties, have similar physical properties to those in the Local Volume especially considering the rough surface brightness limit of our search and their current distance uncertainties.  Right -- The ratio $M_{HI}$/$M_{star}$ as a function of stellar mass for various sample of gas rich galaxies, as well as the UCHVC dwarfs in the current work (represented by color bands).  The SHIELD \citep{Cannon11,McQuinn15} and `Almost Dark' \citep{Cannon15,J15} samples of gas rich galaxies are described in Section~\ref{sec:gas}.  Note the very high $M_{HI}$/$M_{star}$ for ALFALFA-Dw1. \label{fig:properties}}
\end{center}
\end{figure*}

Combining these measurements with our distance estimates presented in Section~\ref{sec:distance}, we can constrain the absolute magnitude and physical half light radii of the new dwarfs (see Table~\ref{table:properties}), and place them into context with respect to other Local Volume dwarfs, as we do in Figure~\ref{fig:properties}.  Absolute $r$-band magnitudes were converted into $V$ band using the SDSS filter transformations of \citet{Jester05}.  Despite their large distance uncertainties (indicated by the colored bands), the new UCHVC dwarf galaxies are similar to other Local Volume dwarfs with ongoing star formation (e.g. dwarf irregulars) in $M_{V}$ versus $r_{h}$ space.  Note that the half light radius for Leo~P has not been measured yet, and so we take the semi-major axis size from \citet{mcquinn13}, which would be an overestimate of the true half light radius.

While a detailed understanding of our dwarf detection limits are difficult to assess given the heterogenous nature of our search, we can make some conservative statements.  First, the Local Group dwarf spheroidals, and the Local Volume dwarf irregulars, clearly suffer from surface brightness selection effects themselves; we have drawn lines of constant central surface brightness (assuming an exponential profile) in Figure~\ref{fig:properties}, and both groups of dwarfs are likely missing ``large and faint" members.   Although our archival search was heterogenous, most of our fields were within the SDSS footprint.  Given this and our search methodology, where the size of any dwarf counterpart had to be $>$5", implies a central surface brightness limit of $\mu_{0}$$\sim$23.5-24 mag arcsec$^{-2}$ for a dwarf with a magnitude of $V$$\sim$17 mag (see Figure~\ref{fig:properties}).  In this sense, it is unsurprising that the majority of the dwarfs uncovered have the properties that they do.  The case of ALFALFA-Dw1 shows that deeper searches may be fruitful, despite the recent deep imaging of B14 and T15 on a subset of the UCHVC sample; there is no trace of this dwarf in the SDSS imaging.


\subsection {Environment}

Even though their distances are unknown, it may still be possible to glean some clues as to the environment that our new UCHVC dwarfs are in. We have searched the NASA Extragalactic Database for all objects on the sky within 3 degrees ($\sim$500 kpc at 10 Mpc or $\sim$170 kpc at 3 Mpc) and 400 km/s of each of our dwarf discoveries. 

A search in the vicinity of ALFALFA-Dw1 (with $v_{LSR}$= $-$123 km s$^{-1}$) turns up many counterparts, as it is projected onto the so-called `Low Velocity Cloud' portion of the Virgo Cluster \citep[at D=17 Mpc; ][]{Gavazzi99}, which has a velocity distribution centered on $v_{LSR}$$\sim$0 km s$^{-1}$ with a dispersion of $\sim$200 km s$^{-1}$ \citep[e.g.][]{Boselli14}.  On the other hand, if the secondary blue clump of stars seen in Figure~\ref{fig:imgs} are really associated with ALFALFA-Dw1, then it is possible that this object is nearby, but with a low overall surface brightness.  Recent observations of Virgo Cluster galaxy IC 3418 have highlighted an extended tail of UV emission and HII regions spanning similar size scales as ALFALFA-Dw1 and its secondary blue clump, likely due to ram pressure stripping \citep{Kenney14}, and it is plausible that we are witnessing something similar in ALFALFA-Dw1 and its secondary clump.  Upcoming HST Cycle 22 observations will possibly pin down its distance and shed light on the nature of ALFALFA-Dw1.

GALFA-Dw2 is possibly associated with the dwarf galaxy UGC 891 ($M_{V}$$=$$-$15), which is $\sim$80' away from it on the sky and has an almost identical velocity ($v_{helio}$=643 km s$^{-1}$). The current best distance estimate for UGC 891 is based on the Tully-Fisher relation, at D=9.4 Mpc \citep{Tully88}.  Assuming that GALFA-Dw2 and UGC 891 are both at this distance, then they are physically separated by $\sim$230 kpc.  In this case, it is possible that GALFA-Dw2 belongs to a UGC 891 `dwarf association'; Leo~P and several of the HI-rich SHIELD galaxies are also in a similar environment \citep{mcquinn13,McQuinn14}.

The remaining three UCHVC dwarfs -- GALFA-Dw1, GALFA-Dw3 and GALFA-Dw4 -- appear to be remarkably isolated, with only a handful of other HI sources (with no known stellar counterpart) turning up in our search.  Similarly isolated objects are rare, but are occasionally reported in the literature \citep[e.g.][]{Pasquali05,Kara14,Kara15}, and present an opportunity to learn about star formation without any external environmental factors.  Again, several of the SHIELD galaxies also appear totally isolated \citep{McQuinn14}, suggesting that the new UCHVC dwarfs come from broadly similar environments. 

\subsection{Gas and Stellar Masses} \label{sec:gas}

HI gas masses, and the gas to stellar mass ratio in particular, tell us about the efficiency of gas conversion to stars in a given galaxy.   We compute HI gas masses (using our assumed distance ranges; Section~\ref{sec:distance}) directly from the data cubes of the GALFA-HI survey for two reasons.  First, a tailored extraction aperture is necessary to capture the entire HI flux from each cloud.  Second, the tabulated values in S12 are in error, and will be corrected in an upcoming GALFA HI update (Saul et al., in prep).  The GALFA-Dw1 and GALFA-Dw2 HI fluxes are unchanged from T15.  As the HI data cubes from A13 are not publicly available, we take the integrated HI flux for ALFALFA-Dw1 directly from that work (integrated HI fluxes using other instrumentation provide consistent results to within $\sim$15\%; E. Adams, private communication).  

To calculate the stellar mass for our UCHVC dwarf sample, we have assumed that ($M_{star}$/$L_{V}$)/($M_{\odot}$/$L_{\odot}$)=1 and translated our absolute V-band magnitudes \citep[with better color measurements of these dwarfs, an improved ($M_{star}$/$L_{V}$)/($M_{\odot}$/$L_{\odot}$) could be assigned, e.g. ][]{Bell03}.  The resulting $M_{HI}$/$M_{star}$ versus $M_{star}$ plot is shown in the right panel of Figure~\ref{fig:properties}, along with similar data for Leo~T \citep{leotgas,leot}, Leo~P,  Local Volume dwarf irregulars \citep[with data from ][]{McConnachie12}, the SHIELD sample of HI-selected low-mass dwarf galaxies \citep{Cannon11,McQuinn15}, and the pilot sample of `Almost Dark' galaxies \citep{Cannon15}.  Several of the new dwarfs have similar $M_{HI}$/$M_{star}$ values as other Local Volume dwarf irregulars, again suggesting that these may just be standard members of this class that have not been uncovered yet.  ALFALFA-Dw1, on the other hand, displays a very large  $M_{HI}$/$M_{star}$$\sim$38, comparable to the `Almost Dark' galaxies \citep{Cannon15,J15}.  This emerging class of galaxies may represent the transition between nearly star-free halos with only HI gas \citep[e.g.][]{Adams15} and those with a $M_{HI}$/$M_{star}$$\sim$1, although these systems still have $>$100 times the stellar mass of objects like Segue~1 \citep{sdssstruct} around the Milky Way (MW).  Nevertheless, they deserve intensive follow up in the ultraviolet, optical and radio regimes.

\section{Conclusions}\label{sec:conclude}

We have presented a systematic, archival UV/optical search for stellar counterparts to the recently uncovered HI UCHVC population, as these systems may shed light on how stars populate dark matter halos in isolated environments.  By searching all available optical and ultraviolet public imaging archives -- along with some supplementary imaging -- we found six compelling dwarf galaxy candidates coincident with HI UCHVC positions.  Spectroscopic followup of all six candidates revealed five with velocities consistent with the HI cloud.  We measured the magnitude and half light radius of all five dwarfs, and their physical properties appear broadly consistent with the faint end of gas rich stellar systems.  

There are many outstanding questions centered on these UCHVC dwarf galaxies. Leading among them are their unknown distances, which will require Hubble Space Telescope imaging, as well as their detailed star formation histories.   Another promising path forward is to understand the HII region metallicities and physical conditions of these dwarfs, which will require deep optical spectroscopy; for instance, recent work on Leo~P has shown it to be among the most metal deficient star forming galaxies ever observed \citep{Skillman13}.

The current sample of UCHVC dwarfs is in some tension with expectations from cosmological simulations.  All of them, with the possible exception of GALFA-Dw1, are likely too distant for comparisons with recent simulations meant to resemble the Local Group, such as ELVIS \citep{garrisonkimmel14}.  What may be more remarkable is the fact that no dwarf counterparts (again, with the possible exception of GALFA-Dw1) were found that could be plausibly associated with the Local Group.  While our search was insensitive to Leo~T-like dwarfs (which is at D$\sim$400 kpc and is resolved into stars by SDSS), we were sensitive to Leo~P-like objects (which is at D$\sim$1.5 Mpc and appears as a blue diffuse object in SDSS). The predicted HI mass function from ELVIS suggests that $\sim$50 unidentified dwarf galaxies with $M_{HI}>10^5 M_{\odot}$ may be within D$\sim$1.2 Mpc of the MW or M31 \citep{garrisonkimmel14}, which is comparable to the $M_{HI}$ detection limits at that distance for both GALFA-HI and ALFALFA.  Given the sky coverage of the GALFA-HI Data Release 1 \citep[18\% of the sky;][]{Peek11} and ALFALFA $\alpha$.40 Data Release \citep[$\sim$7\% of the sky;][]{Haynes11}, $\sim$5-10 dwarf counterparts would thus be expected.  Either these potential galaxies a) never formed stars and so would be invisible to our search; b) are too near to us and so would appear as resolved stars rather than diffuse galaxies in our search (and so would have gone undiscovered) or c) there is a conflict with expectations from recent $\Lambda$CDM simulations.  Clearly, distance measurements to the newly discovered dwarfs, and deeper searches for stellar counterparts to the other UCHVCs, are necessary before decisive conclusions can be drawn.

Finally, our archival search for UV/optical counterparts uncovered four dwarf galaxies associated with the GALFA-HI Compact Cloud Catalog (consisting of 27 sources overall), while the ALFALFA UCHVC catalog (consisting of 59 total sources) uncovered only a single object.  Some effort was made by the ALFALFA team to not include UCHVCs which had clear optical counterparts in DSS/SDSS into the final UCHVC catalog of A13, so this may be the root of the difference between the two surveys (E. Adams, private communication).  Once the GALFA-HI fluxes have been amended, it will be important to do a cross-comparison between the ALFALFA and GALFA UCHVC catalogs, and identify which physical HI-related properties are most important for yielding a dwarf galaxy companion.  A broader search of compact HI clouds may be warranted in order to uncover faint dwarf galaxies that may be at the edge of the selection criteria that are currently being employed by the different HI teams.  Another path forward is to search the SDSS archives for blue diffuse galaxies \citep[e.g.][]{James14}, or to examine the GALEX archive.  In any case, as ALFALFA-Dw1 has shown, there may yet be several interesting stellar systems associated with the UCHVC population, but below the detection limit of surveys such as SDSS and DSS; further deep followup work is needed.

\acknowledgments
We warmly thank Maureen Conroy, John Roll and Sean Moran for their prolonged efforts and help related
to Magellan/Megacam.  We thank Alison Marqusee, Chris Nagele, and Eric Smith for assistance obtaining ARC 3.5m observations.  We are also grateful to Elizabeth Adams for useful comments. DJS, JDS and PG acknowledge support from
NSF grant AST-1412504.  BW and JH acknowledge support from NSF AST-1151462.  PG acknowledges additional
support from NSF grant AST-1010039. EO thanks the NSF for support through AST-1313006.
The Digitized Sky Surveys were produced at the Space Telescope Science Institute under U.S. Government grant NAG W-2166. The images of these surveys are based on photographic data obtained using the Oschin Schmidt Telescope on Palomar Mountain and the UK Schmidt Telescope. The plates were processed into the present compressed digital form with the permission of these institutions.
This research has made use of the NASA/IPAC Extragalactic Database (NED) which is operated by the Jet Propulsion Laboratory, California Institute of Technology, under contract with the National Aeronautics and Space Administration.
Funding for SDSS-III has been provided by the Alfred P. Sloan Foundation, the Participating Institutions, the National Science Foundation, and the U.S. Department of Energy Office of Science. The SDSS-III web site is http://www.sdss3.org/.
Based in part on data collected at the Subaru Telescope and obtained from SMOKA, which is operated by the Astronomy Data Center, National Astronomical Observatory of Japan.
This research was made possible through the use of the AAVSO Photometric All-Sky Survey (APASS), funded by the Robert Martin Ayers Sciences Fund.
Observations reported here were obtained at the MMT Observatory, a joint facility of the University of Arizona and the Smithsonian Institution.
This paper includes data gathered with the 6.5 meter Magellan Telescopes located at Las Campanas Observatory, Chile.
Based on observations obtained at the Southern Astrophysical Research (SOAR) telescope, which is a joint project of the Minist\'{e}rio da Ci\^{e}ncia, Tecnologia, e Inova\c{c}\~{a}o (MCTI) da Rep\'{u}blica Federativa do Brasil, the U.S. National Optical Astronomy Observatory (NOAO), the University of North Carolina at Chapel Hill (UNC), and Michigan State University (MSU).

\appendix

\section{UCHVC Clarifications}\label{sec:clarifications}

A handful of the UCHVC sources have already appeared in the literature or are coincident with known galaxies.  We briefly describe each case here.

{\it GALFA 147.0+07.1+525 -- } This GALFA nearby galaxy candidate is also in the $\alpha$.40 catalog, listed as AGC~191803 \citep{Haynes11}.  Coincident with the HI source is a blue, disky galaxy, easily detectable in SDSS bands and in GALEX archival data.  It also is listed in the Updated Nearby Galaxy Catalog \citep{K13}, which reports a Tully-Fisher distance of 14.9 Mpc, whereas the Hubble Flow distance is 7.6 Mpc.  While it is likely a formality, to our knowledge there is no H$\alpha$ velocity confirming the association between the HI source and the blue, coincident galaxy.  As this galaxy has been reported elsewhere, we do not include it in our main sample of new discoveries, although further followup is warranted.

{\it GALFA 188.9+14.5+387 -- }  This GALFA nearby galaxy candidate appears coincident with M91, albeit at a somewhat lower velocity ($v_{LSR}$=403.41 km s$^{-1}$ with FWHM = 68.76 km s$^{-1}$; versus $v_{LSR}$=491 km s$^{-1}$ for M91).   As pointed out by \citet{GrcevichPhD}, it is plausible that this gas cloud is an extension of M91's halo or that it is a projected dwarf galaxy.

{\it GALFA 187.5+08.0+473 -- } This gas cloud is projected between VCC~1249 and M49, and has been previously identified by \citet{McNamara94}.  It is almost certainly stripped gas from a past encounter between VCC~1249 and M49, as has been discuss at length by \citet{Battaia12}.

{\it ALFALFA 320.95+72.32+185 -- } This gas cloud is projected 1.4' away from UGC 08298, an irregular galaxy with v$_{\odot}$=1157 km s$^{-1}$.  It is unlikely that the ALFALFA source, at $v_{LSR}$=185  km s$^{-1}$, is  associated with UGC08298, given the large velocity difference between the two systems.

\bibliographystyle{apj}
\bibliography{mybib}

\clearpage

\clearpage

\clearpage
\LongTables
\tabletypesize{\scriptsize}

\begin{deluxetable*}{lcccccc}
\tablecolumns{7}
\tablecaption{Summary of UCHVC Archival Search \label{table:archive}}
\tablehead{
\colhead{Survey} & \colhead{RA} & \colhead{DEC} &  \colhead{HI $v_{LSR}$} & \colhead{Optical} & \colhead{GALEX} & \colhead{Comments}\\
\colhead{ID} & \colhead{(J2000)} & \colhead{(J2000)}& \colhead{(km s$^{-1}$)}& \colhead{Archives\tablenotemark{a}} & \colhead{Archive} & \colhead{}}
\\
\startdata
\hline
\multicolumn{7}{c}{GALFA-HI Survey}\\
\hline
003.7+10.8+236 &  00:14:45 & +10:49:00 & 235.38 & SDSS & NUV\tablenotemark{b} & GALFA-Dw1\tablenotemark{$*$}\\
019.8+11.1+617 & 01:19:13 & +11:07:00 & 611.63 & SDSS & NUV, FUV\tablenotemark{b} & GALFA-Dw2\tablenotemark{$*$}\\
044.7+13.6+528  & 02:58:57 & +13:37:00 & 528.59 & DSS\tablenotemark{d} & NUV, FUV\tablenotemark{b} & GALFA-Dw3\tablenotemark{$*$}\\
063.7+33.3+447 & 04:14:57 & +33:18:00 & 447.77 & DSS & NUV, FUV\tablenotemark{b} & \\
084.4+24.0+152 & 05:37:33 & +24:02:00 & 153.04 & DSS & NUV\tablenotemark{b} & \\
086.4+31.8+612 & 05:45:45 & +31:46:00 & 612.27 & DSS & ... & \\
086.4+10.8+611  & 05:45:45 & +10:46:00 & 614.53 & DSS\tablenotemark{d} & ... & GALFA-Dw4\tablenotemark{$*$}\\
090.9+30.9+266 & 06:03:37 & +30:53:00 & 267.56 & DSS & NUV\tablenotemark{b} & \\
092.1+09.5+584 & 06:08:33 & +09:29:00 & 584.58 & CFHT(r)\tablenotemark{e} & ... & \\
100.0+36.7+417 & 06:39:49 & +36:41:00 & 416.43 & SDSS & NUV,FUV\tablenotemark{b} & \\
100.9+09.2+311 & 06:43:33 & +09:13:00 & 310.11 & SDSS & ... & \\
104.4+04.1$-$145 & 06:57:37 & +04:07:00 & $-$157.20 & DSS & ...  & \\
143.7+12.9+223 & 09:34:41 & +12:51:00 & 222.93 & SDSS& NUV\tablenotemark{b} & \\
147.0+07.1+525 & 09:48:05 & +07:08:00 & 526.31 & SDSS & NUV, FUV\tablenotemark{c} & see Appendix \\
162.1+12.5+434  & 10:48:25 & +12:31:00 & 435.67 & SDSS,CFHT(i)\tablenotemark{g} & NUV, FUV\tablenotemark{f} & counterpart with $\Delta$v$\sim$5260 km s$^{-1}$ \\
183.0+04.4$-$112 & 12:12:10 & +04:23:00 & $-$112.84 & SDSS & NUV\tablenotemark{v} & \\
184.8+05.7$-$092 & 12:19:06 & +05:40:00 & $-$89.90 & SDSS,CFHT(u,g,i)\tablenotemark{h} & NUV, FUV\tablenotemark{w} & \\
187.5+08.0+473 & 12:29:54 & +07:58:00 & 473.70 & SDSS, CFHT(u,g,i,z)\tablenotemark{i} & NUV, FUV\tablenotemark{b} & see Appendix\\
188.9+14.5+387 & 12:35:26 & +14:30:00 & 403.41 & SDSS & NUV, FUV\tablenotemark{b} & see Appendix\\
195.9+06.9$-$100 & 13:03:38 & +06:55:00 & $-$104.44 & SDSS & NUV,FUV\tablenotemark{b} & \\
196.6+06.5$-$105 & 13:06:18 & +06:29:00 & $-$103.55 & SDSS & NUV\tablenotemark{x} & \\
215.9+04.6+205 & 14:23:26 & +04:33:00 & 217.30 & SDSS & NUV, FUV\tablenotemark{c} & HVC351.17+58.56+214\\
331.8+21.0+303 & 22:07:06 & +20:59:00 & 303.48 & SDSS,CFHT(r)\tablenotemark{j} & ... & \\
339.0+09.0$-$237 & 22:36:06 & +09:02:00 & $-$236.64 & SDSS & NUV\tablenotemark{c} &\\
341.7+07.7$-$234 & 22:46:58 & +07:41:00 & $-$234.09 & SDSS & ... & \\
342.1+20.6+208 & 22:48:26 & +20:33:00 & 208.14 & SDSS & NUV\tablenotemark{c}, FUV\tablenotemark{b} & \\
345.0+07.0$-$245 & 22:59:58 & +07:01:00 & $-$244.33 & SDSS & NUV, FUV\tablenotemark{b} &\\
\hline
\multicolumn{7}{c}{ALFALFA Survey}\\
\hline
111.65$-$30.53$-$124 & 00:05:54.3 & +31:20:14 & $-$124 & SDSS & NUV\tablenotemark{b} &\\
123.11$-$33.67$-$176 & 00:52:06.2 & +29:12:04 & $-$176 & SDSS & NUV,FUV\tablenotemark{b} &   \\
123.74$-$33.47$-$289 & 00:54:31.6 & +29:24:02 & $-$289 & SDSS & NUV,FUV\tablenotemark{b} &  \\
126.85$-$46.66$-$310 & 01:02:37.8 & +16:07:52 & $-$310 & SDSS & NUV,FUV\tablenotemark{b} &  \\
131.90$-$46.50$-$276 & 01 17 03.4 & +15:55:48 & $-$276 & SDSS\tablenotemark{d} & NUV,FUV\tablenotemark{b} &  \\
137.90$-$31.73$-$327 & 01:49:52.1 & +29:26:00 & $-$327 & SDSS,CFHT(r)\tablenotemark{k} & NUV,FUV\tablenotemark{b} \\
138.39$-$32.71$-$320 & 01:50:31.4 & +28:22:59 & $-$320 & SDSS,CFHT(r)\tablenotemark{l} & NUV,FUV\tablenotemark{b}\\
154.00$-$29.03$-$141 & 02:52:29.7 & +26:26:30 & $-$141 & DSS & ...\\
196.50+24.42+146 & 07:55:27.1 & +24:41:43 & 146 & SDSS & NUV,FUV\tablenotemark{b} \\
196.09+24.74+166 & 07:56:14.8 & +25:09:00 & 166 & SDSS & NUV,FUV\tablenotemark{c} & B14 \\
198.48+31.09+165 & 08:25:46.7 & +25:11:28 & 165 & SDSS & NUV,FUV\tablenotemark{b} \\ 
204.88+44.86+147 & 09:30:13.2 & +24:12:17 & 147 & SDSS & NUV,FUV\tablenotemark{b} & B14\\
205.28+18.70+150 & 07:45:59.9 & +14:58:37 & 150 & SDSS,CFHT(r)\tablenotemark{m} & NUV,FUV\tablenotemark{b} & B14\\ 
234.33+51.28+143 & 10:27:01.1 & +08:47:08 & 143 & SDSS,CFHT(g)\tablenotemark{n} & ... \\
252.98+60.17+142 & 11:21:19.6 & +06:21:32 & 142 & SDSS & ... \\
245.26+69.53+217 & 11:40:08.1 & +15:06:44 & 217 & SDSS & ... & B14\\
255.76+61.49+181 & 11:28:55.6 & +06:25:29 & 181 & SDSS & NUV,FUV\tablenotemark{b} & B14\\
274.68+74.70$-$123 & 12:21:54.7 & +13:28:10 & $-$123 & SDSS\tablenotemark{d},CFHT(u,g,i,z)\tablenotemark{i} & NUV\tablenotemark{y} & ALFALFA-Dw1; B14 \\
290.19+70.86+204 & 12:34:40.2 & +08:24:08 & 204 & SDSS,CFHT(u,g,r,i,z)\tablenotemark{o} & NUV,FUV\tablenotemark{b} & B14 \\
292.94+70.42+159 & 12:37:58.5 & +07:48:49 & 159 & SDSS,CFHT(g,r,i,z)\tablenotemark{p} & NUV\tablenotemark{z} & \\
028.09+71.86-144 & 14:10:58.1 & +24:12:04 & $-$144 & SDSS & NUV,FUV\tablenotemark{b} & B14\\
253.04+61.98+148 & 11:26:24.8 & +07:39:15 & 148 & SDSS & NUV,FUV\tablenotemark{b} \\
256.34+61.37+166 & 11:29:28.6 & +06:09:23 & 166 & SDSS & NUV,FUV\tablenotemark{c} \\
250.16+57.45+139 & 11:09:29.8 & +05:26:01 & 139 & SDSS & NUV,FUV\tablenotemark{c} & B14\\
277.25+65.14$-$140 & 12:09:20.0 & +04:23:30 & $-$140 & SDSS & NUV\tablenotemark{c} & B14\\
295.19+72.63+225 & 12:42:04.6 & +09:54:05 & 225 & SDSS,CFHT(g,r,i,z)\tablenotemark{q} & NUV\tablenotemark{aa}\\
298.95+68.17+270 & 12:45:29.8 & +05:20:23 & 270 & SDSS & NUV\tablenotemark{c} & B14\\
320.95+72.32+185 & 13:13:21.5 & +10:12:57 & 185 & SDSS & NUV\tablenotemark{b} & see Appendix\\
324.03+75.51+135 & 13:12:42.3 & +13:30:46 & 135 & SDSS & NUV,FUV\tablenotemark{b} & B14\\
326.91+65.25+316 & 13:30:43.8 & +04:13:38 & 316 & SDSS & NUV\tablenotemark{c} & B14\\
330.13+73.07+132 & 13:22:41.6 & +11:52:31 & 132 & SDSS & NUV,FUV\tablenotemark{b} & B14\\
351.17+58.56+214 & 14:23:21.2 & +04:34:37 & 214 & SDSS\tablenotemark{d} & NUV,FUV\tablenotemark{b} & GALFA 215.9+04.6+205; B14\\
352.45+59.06+263 & 14:23:57.7 & +05:23:40 & 263 & SDSS & ... & B14\\
353.41+61.07+257 & 14:19:48.6 & +07:11:15 & 257 & SDSS & NUV,FUV\tablenotemark{b} & B14\\
356.81+58.51+148 & 14:31:58.8 & +06:35:20 & 148 & SDSS & NUV,FUV\tablenotemark{b} & B14\\
005.58+52.07+163 & 15:04:41.3 & +06:12:59 & 163 & SDSS & NUV,FUV\tablenotemark{c} & B14\\
013.59+54.52+169 & 15:07:23.0 & +11:32:56 & 169 & SDSS & NUV,FUV\tablenotemark{b} & B14\\
013.60+54.23+179 & 15:08:24.4 & +11:24:22 & 179 & SDSS & NUV,FUV\tablenotemark{b}\tablenotemark{c} & B14\\
013.63+53.78+222 & 15:10:00.6 & +11:11:27 & 222 & SDSS & NUV,FUV\tablenotemark{c} & B14\\
026.01+45.52+161 & 15:55:07.5 & +14:29:29 & 161 & SDSS & NUV,FUV\tablenotemark{ab} & B14\\
026.11+45.88+163 & 15:53:54.0 & +14:41:48 & 163 & SDSS & NUV,FUV\tablenotemark{ab} \\
028.07+43.42+150 & 16:05:32.6 & +14:59:20 & 150 & SDSS & NUV,FUV\tablenotemark{ac} & B14\\
028.47+43.13+177 & 16:07:07.0 & +15:08:31 & 177 & SDSS & NUV,FUV\tablenotemark{b} \\
029.55+43.88+175 & 16:05:29.4 & +16:09:12 & 175 & SDSS & NUV,FUV\tablenotemark{b}\\
028.03+41.54+127 & 16:12:36.8 & +14:12:27 & 127 & SDSS & NUV,FUV\tablenotemark{b}\\
028.66+40.38+125 & 16:17:45.3 & +14:10:36 & 125 & SDSS & NUV,FUV\tablenotemark{b}\\
019.13+35.24-123 & 16:22:35.7 & +05:08:48 & $-$123 & SDSS & NUV,FUV\tablenotemark{b}\\
027.86+38.25+124 & 16:24:43.4 & +12:44:12 & 207 & SDSS & NUV,FUV\tablenotemark{c} & B14\\
080.69$-$23.84$-$334 & 22:01:00.7 & +24:44:04 & $-$334 & SDSS & NUV,FUV\tablenotemark{b}\\
082.91$-$20.46$-$426 & 21:58:02.9 & +28:37:35 & $-$426 & SDSS & NUV,FUV\tablenotemark{b}\\
082.91$-$25.55$-$291 & 22:12:38.6 & +24:43:11 & $-$291 & SDSS,CFHT(r)\tablenotemark{r} & NUV,FUV\tablenotemark{b}\\
084.01$-$17.95$-$311 & 21:54:06.2 & +31:12:49 & $-$311 & DSS & NUV\tablenotemark{c},FUV\tablenotemark{b}\\
084.61$-$26.89$-$330 & 22:21:34.4 & +24:36:38 & $-$330 & SDSS & NUV,FUV\tablenotemark{b}\\
086.18$-$21.32$-$277 & 22:11:21.8 & +29:54:02 & $-$277 & SDSS,CFHT(r)\tablenotemark{s}& NUV,FUV\tablenotemark{b}\\
087.35$-$39.78$-$434 & 23:00:56.4 & +15:20:14 & $-$454 & SDSS & ... & \\
088.15$-$39.37$-$445 & 23:02:11.3 & +16:00:48 & $-$445 & SDSS,CFHT(u,g,r,i)\tablenotemark{t} & ... & \\
092.53$-$23.02$-$311 & 22:38:23.4 & +31:52:57 & $-$311 & SDSS & NUV,FUV\tablenotemark{b}\\
108.98$-$31.85$-$328 & 23:56:58.8 & +29:32:35 & $-$328 & SDSS & NUV\tablenotemark{b}\\
109.07$-$31.59$-$324 & 23:57:02.1 & +29:48:46 & $-$324 & SDSS & NUV\tablenotemark{b}\\
\enddata
\tablecomments{Fields with `B14' noted in the comments are fields with deep imaging reported by \citet{Bellazzini14}.  None of these found dwarf counterparts, besides ALFALFA-Dw1\\
$^*$Candidate diffuse dwarf galaxy at UCHVC position. GALFA-Dw1 and GALFA-Dw2 are Pisces~A and Pisces~B, as reported by T15.  ALFALFA-Dw1 is SECCO 1 as reported by \citet{Bellazzini15}. $^a$DSS images are listed only when it is the only option. $^b$From the GALEX All Sky Imaging Survey; $t_{exp}$$\sim$100s of seconds. $^c$From the GALEX Medium Imaging Survey; $t_{exp}$$\sim$1500 s. $^d$Supplementary imaging taken; see Section~\ref{sec:imaging} and Table~\ref{table:imaging}. $^e$CFHT; r $t_{exp}$=150 s. $^f$GALEX Guest Investigator data; NUV,FUV $t_{exp}$=1667 s. $^g$CFHT i-band $t_{exp}$=476 s. $^h$CFHT (u,g,i) $t_{exp}$=(2400, 2760, 3240) s. $^i$CFHT (u, g, i, z) $t_{exp}$=(6402, 3170, 2055, 4400) s. $^j$ CFHT r $t_{exp}$=420 s. $^k$CFHT r $t_{exp}$= 1600 s. $^l$CFHT r $t_{exp}$= 2260 s. $^m$CFHT r $t_{exp}$= 200 s. $^n$CFHT g $t_{exp}$= 1050 s. $^o$CFHT (u,g,r,i,z) $t_{exp}$=(6800, 3100, 3720, 6000, 7500) s. $^p$CFHT (g, r, i, z) $t_{exp}$=(3170, 4461, 2055, 4400) s. $^q$CFHT (g,r,i,z) $t_{exp}$=(4800, 1400, 7500, 4400) s. $^s$CFHT r $t_{exp}$= 480 s. $^t$CFHT r $t_{exp}$= 240 s, $^u$CFHT (u,g,r,i) $t_{exp}$=(2800, 5400, 920, 470) s, $^v$GALEX Guest Investigator data; NUV $t_{exp}$=1730 s. $^w$GALEX Guest Investigator data; FUV,NUV $t_{exp}$=1655, 1655 s, $^x$GALEX Guest Investigator data; NUV $t_{exp}$=1650 s., $^y$GALEX Guest Investigator data; NUV $t_{exp}$=1690 s.$^z$GALEX Guest Investigator data; NUV $t_{exp}$=1605 s.$^{aa}$GALEX Guest Investigator data; NUV $t_{exp}$=1909 s.$^{ab}$GALEX Guest Investigator data; FUV, NUV $t_{exp}$=1500, 1500 s.$^{ac}$GALEX Guest Investigator data; FUV, NUV $t_{exp}$=5057, 5057 s.}
\end{deluxetable*}

\clearpage

\begin{deluxetable*}{lcccccccccc}
\tablecolumns{7}
\tablecaption{Supplementary Imaging Log \label{table:imaging}}
\tablehead{
\colhead{Target}  & \colhead{Telescope/} &\colhead{UT Date} & \colhead{Filter} & \colhead{Exposure} & \colhead{Depth} & \colhead{Notes} \\
& \colhead{Instrument} & & &(sec) & (mag)
}\\
\startdata
HVC274.68+74.70$-$123 & Magellan/Megacam & 11 June 2013 & g & 8$\times$300 & 25.0 & ALFALFA-Dw1\\
& & & r & 6$\times$300 & 25.1 \\
HVC351.17+58.56+214\tablenotemark{b}& Magellan/Megacam & 27 April 2014 & g & 6$\times$300 & 25.9\\
& & & r& 6$\times$300 & 25.9 \\
GALFA 044.7+13.6+528 & Magellan/Megacam & 26 Oct 2014 & g & 7$\times$300 & 25.3  & GALFA-Dw3\\
& & & r & 8$\times$300 & 25.5 \\
HVC131.90$-$46.50$-$276 & APO/SPICAM & 17 Nov 2014 & g & 2$\times$540 & 23.3\\
& & & r & 2$\times$540 & 23.5 \\
GALFA 086.4+10.8+611 & APO/SPICAM & 17 Nov 2014 & g & 4$\times$450 & 23.6 & GALFA-Dw4 \\
& & & r & 4$\times$450 & 23.8 & 
\enddata
\tablenotetext{b}{Also identified as GALFA 215.9+04.6+205} 

\end{deluxetable*}

\clearpage

\begin{deluxetable*}{lcccccccccc}
\tablecolumns{6}
\tablecaption{Spectroscopy Log \label{table:spec}}
\tablehead{
\colhead{Target}  & \colhead{UCHVC} &\colhead{Telescope/} &\colhead{UT Date}   & \colhead{Exposure} & \colhead{H${\alpha}$ $v_{LSR}$} &  \colhead{Notes}\\
& \colhead{ID} & \colhead{Instrument} & & (sec) & (km s$^{-1}$)
}\\
\startdata
ALFALFA-Dw1 & HVC274.68+74.70$-$123 & Magellan/IMACS & 17 June 2014 &  5$\times$1800 & $-$114$\pm$12 & SECCO 1 \\
GALFA-Dw1 & GALFA 003.7+10.8+236 & SOAR/Goodman & 29 June 2014 & 5$\times$900 & 236$\pm$8 & Pisces~A\\ 
GALFA-Dw2 & GALFA 019.8+11.1+617 & MMT/BCS & 21 Oct 2014 &  1$\times$900  & 625$\pm$30 & Pisces~B\\
GALFA-Dw3 & GALFA 044.7+13.6+528 & APO/DIS & 18 Nov 2014 & 5$\times$1200 & 503$\pm$35 \\
GALFA-Dw4 & GALFA 086.4+10.8+611& APO/DIS & 18 Nov 2014 & 3$\times$1200 & 607$\pm$35 \\
SDSS J1048+1230 & GALFA 162.1+12.5+434 & MMT/BCS & 24 Dec 2014 & 2$\times$900 & 5730$\pm$200 & Not ass. with UCHVC
\enddata

\end{deluxetable*}

\clearpage

\begin{deluxetable*}{lcccccccccccccc}

\tablecolumns{6}
\tablecaption{Properties of the New Dwarf Galaxies \label{table:properties}}
\tablehead{
\colhead{Dwarf}  & GALFA-Dw1\tablenotemark{a} & GALFA-Dw2\tablenotemark{b} & GALFA-Dw3 & GALFA-Dw4 & ALFALFA-Dw1\tablenotemark{c} \\
 }
\\
\startdata
Survey ID & 003.7+10.8+236 & 019.8+11.1+617 & 044.7+13.6+528 & 086.4+10.8+611 & 274.68+74.70$-$123\\
RA (J2000) (optical) & 00:14:46.09 & 01:19:11.72 & 02:58:56.24 & 05:45:44.79 & 12:21:54.02 \\
DEC (J2000) (optical) & +10:48:47.1&  +11:07:16.3 & +13:37:47.9 & +10:46:15.6 &+13:27:37.32 \\
l ($^{\circ}$) & 108.52 & 133.83 & 164.14 & 195.66 & 274.68\\
b ($^{\circ}$) & -51.03 & -51.16 & -38.83 & -9.32 & 74.69\\
E(B-V) (mag) & 0.09 & 0.05 & 0.13 & 0.54 & 0.05\\
HI $v_{LSR}$ (km s$^{-1}$) & 235.38 & 611.63 & 528.59 & 614.53 & $-$123\\
$H{\alpha}$ $v_{LSR}$ (km s$^{-1}$) & 236$\pm$8 & 625$\pm$30 & 503$\pm$35 & 607$\pm$35 & $-$114$\pm$12\\
$r_{0}$ (mag) & 17.12$\pm$0.19 & 16.91$\pm$0.13 & 16.43$\pm$0.32 & 14.52$\pm$0.08 & 20.51$\pm$0.26 \\
$(g-r)_{0}$ (mag) & 0.19$\pm$0.22 & 0.09$\pm$0.16 & 0.32$\pm$0.32 & 0.06$\pm$0.14 & $-$0.10$\pm$0.35\\
$NUV_{0}$ (mag) & 18.48$\pm$0.15 & 18.50$\pm$0.19 & 17.58$\pm$0.60 & ... & 20.73$\pm$0.70\\
$FUV_{0}$ (mag) & ... & 18.64$\pm$0.44 & 17.57$\pm$0.38 & ... & ...\\
$r_{h}$ (arcsec) & 8.6$\pm$1.8 & 10.2$\pm$2.6 & 9.2$\pm$1.5 & 12.5$\pm$0.6 & 8.1$\pm$0.9\\
$F_{HI}$ (Jy km s$^{-1}$) & 1.22$\pm$0.07\tablenotemark{d} & 1.6$\pm$0.2\tablenotemark{d} & 1.7$\pm$0.1 & 1.89$\pm$0.25 &0.92\tablenotemark{e} \\
\hline
\hline
\multicolumn{6}{c}{Distance Dependent Properties}\\
\hline
\hline
Est. Dist range (Mpc) & 1.7 -- 5.3 & 3.5 -- 13.4 & 3.0 -- 11.4 & 3.0 -- 13.3 & 3.0 -- 18.0 \\
$M_{r}$ (mag) & $-$9.0 to $-$11.5& $-$10.8 to $-$13.7 & $-$10.9 to $-$13.8 & $-$12.9 to $-$16.1 & $-$6.9 to $-$10.8 \\
$r_{h}$ (pc) & 71 -- 220 & 173 -- 663 & 133 -- 510 & 181 -- 806 & 117 -- 707 \\
$M_{*}$ (10$^{5}$ $M_{\odot}$) & 3.3 -- 33.0 & 18.1 -- 265.6 & 19.0 -- 276.2 & 121.7 -- 2392.0 & 0.5 -- 18.7 \\
$M_{HI}$ (10$^{5}$ $M_{\odot}$) & 8.3 -- 80.6 & 46.2 -- 677.9 & 36.0 -- 524.3 & 40.0 -- 787.8 & 19.5 -- 702.3
\enddata
\tablenotetext{a}{Alternatively named Pisces~A by T15}
\tablenotetext{b}{Alternatively named Pisces~B by T15}
\tablenotetext{c}{Alternatively named SECCO 1 by B15}
\tablenotetext{d}{HI flux as reported in T15}
\tablenotetext{e}{HI flux as reported in A13; no uncertainty presented}
\end{deluxetable*}

\end{document}